\documentclass[useAMS, usenatbib]{mn2e}
\usepackage{graphicx}
\usepackage[footnotesize, bf]{caption}
\usepackage{color}
\usepackage{natbib}
\usepackage{tabularx}
\usepackage{rotating}
\usepackage{url}
\usepackage{longtable}
\usepackage{supertabular}
\usepackage{multirow}

\begin{document}
\title[Line contamination in the SCUBA-2 continuum data]
	{Molecular line contamination in the SCUBA-2 450~$\micron$ and 850~$\micron$ continuum data}
\author[E. Drabek et al.]{E.~Drabek$^1$\thanks{E-mail:  emily@astro.ex.ac.uk}, J.~Hatchell$^1$,  P.~Friberg$^2$, J. Richer$^{3, 4}$,  S. Graves$^{3, 4}$, J. V. Buckle$^{3, 4}$, 
\newauthor D. Nutter$^{5}$, D. Johnstone$^{6, 7}$, J. Di Francesco$^{6, 7}$ \\
$^1$University of Exeter, School of Physics, Stocker Rd, Exeter EX4 4QL, United Kingdom \\
$^2$Joint Astronomy Centre, 660 North A'ohoku Place, Hilo, HI 96720, USA \\ 
$^3$Astrophysics Group, Cavendish Laboratory, J J Thomson Avenue, Cambridge CB3 0HE \\
$^4$Kavli Institute for Cosmology, Cambridge, Madingley Road, Cambridge, CB3 0HA \\
$^5$School of Physics \& Astronomy, Cardiff University, 5 The Parade, Cardiff, UK \\
$^6$National Research Council Canada, Herzberg Institute of Astrophysics, 5071 West Saanich Rd, Victoria, BC, V9E 2E7 \\
$^7$Department of Physics \& Astronomy, University of Victoria, 3800 Finnerty Rd., Victoria, BC, Canada \\
}

\maketitle
\label{firstpage}

\begin{abstract}
Observations of the dust emission using millimetre/submillimetre bolometer arrays can be contaminated by molecular line flux, such as flux from $^{12}$CO. As the brightest molecular line in the submillimetre, it is important to quantify the contribution of CO flux to the dust continuum bands.  
%the continuum flux in order to distinguish between observations of fainter objects with higher sensitivity detectors and contamination to the dust continuum bands due to molecular lines. 
Conversion factors were used to convert molecular line integrated intensities to flux detected by bolometer arrays in mJy~beam$^{-1}$.  These factors were calculated  for $^{12}$CO line integrated intensities to the SCUBA-2 850~$\micron$ and 450~$\micron$ bands.  The conversion factors were then applied to HARP $^{12}$CO~3--2 maps of NGC~1333 in the Perseus complex and NGC~2071 and NGC~2024 in the Orion~B molecular cloud complex to quantify the respective $^{12}$CO flux contribution to the 850~$\micron$ dust continuum emission.  Sources with high molecular line contamination were analysed in further detail for molecular outflows and heating by nearby stars to determine the cause of the $^{12}$CO contribution.  The majority of sources had a $^{12}$CO 3--2 flux contribution under 20~per~cent.  However, in regions of molecular outflows, the $^{12}$CO can dominate the source dust continuum (up to 79~per~cent contamination) with $^{12}$CO fluxes reaching $\sim$68~mJy~beam$^{-1}$.  %of sources with a dust continuum flux ranging from 100-1160 mJy, particularly in high continuum flux regions ($>$ 550 mJy).  Hot ambient gas resulting from nearby stars can also be a cause for contamination, but only in lower flux sources ($<$ 550 mJy).  
\end{abstract}

\section{Introduction}
\label{introduction}  

Dust continuum is a useful tracer of star formation.   
%The classification of young stellar objects (YSOs) are determined based on their spectral energy distributions (SEDs), calculated using the spectral index $\alpha$ in the 1-10 $\mu$m (near to mid infrared) range, which is the gradient of the flux vs wavelength (Lada 1987). 
The submillimetre (roughly 300 to 1000~$\micron$) detection of the dust emission identifies Class 0 and prestellar cores as well as disks, filamentary structure in molecular clouds, and the dust and gas masses of galaxies.  In order to quantify the flux from dust in the submillimetre wavelengths, the heat generated by the radiation is measured by bolometers with the detected wavelength range defined by wide-band filters %bolometers measure changes in heat input from observations and converts these to a measurable quantity (voltage or current) over a specific bandwidth 
\citep{2002ASPC..278..463H}. These observations of broadband continuum emission from the dust can be contaminated by molecular line flux, particularly from $^{12}$CO, which is the second most abundant molecule in the interstellar medium (after H$_2$) with strong emission lines in the submillimetre \citep{2003A&A...412..157J, 2004MNRAS.349.1428S, 2009A&A...502..139H, 2003ApJ...588..243Z}.  Since the molecular line contamination depends explicitly on the bandwidth and wavelength of the bolometer, it is important to quantify the potential contribution from molecular lines to make accurate flux measurements of the submillimetre dust emission used to calculate masses.  

The CO line contribution can be quantified by comparing observations of the dust continuum emission and the CO line emission \citep{1995A&A...301..853G}. Past research \citep{1999ApJ...510L..49J, 2000MNRAS.318..952D, 2000ApJ...537..631P, 2002ApJ...580..285T} has studied the Submillimetre Common User Bolometer Array (SCUBA) at the James Clerk Maxwell Telescope (JCMT), where line contribution from the $^{12}$CO~3--2~line was found to range from little to tens of per~cent in the 850~$\micron$ band.  Other studies have examined contamination in various bolometer instruments, including MAMBO-II, Bolocam, and SHARC-II.  The Submillimetre High Angular Resolution Camera~II (SHARC-II) operates at the same wavelength range as SCUBA (450~$\micron$ and 850~$\micron$), but also includes a 350~$\micron$ filter (780 to 910~GHz).  The 350~$\micron$ SHARC-II  continuum could be potentially contaminated by the $^{12}$CO~7--6~line (806~GHz) up to $\sim$20~per~cent, similar to the $^{12}$CO~3--2 contamination to the SCUBA 850~$\micron$ continuum \citep{2009A&A...502..139H}.  While Bolocam (operated at Caltech at 1.1~mm with a 250 to 300~GHz filter) has been designed to exclude $^{12}$CO line contamination, the Max Planck Millimetre Bolometer~II (MAMBO-II; operated by the Max Planck Institut f\"{u}r Radioastronomie at 1.2~mm with a $\sim$210 to 290~GHz filter) includes $^{12}$CO~2--1 (230~GHz) molecular line emission which could potentially increase flux at most a few percent \citep{2011ApJS..192....4A}.  %Neither Bolocam (operated at Caltech at 1.1~mm with a 250 to 300~GHz filter) nor the Max Planck Millimetre Bolometer~II (MAMBO-II; operated by the Max Planck Institut f\"{u}r Radioastronomie at 1.2~mm with a $\sim$210 to 290~GHz filter) include $^{12}$CO molecular line emission.  
Other possible contamination for Bolocam and MAMBO-II could result from other known molecular lines in clouds, including SiO~6--5 ($\sim$260~GHz) and HCN~3--2 ($\sim$258~GHz).

The successor to SCUBA is SCUBA-2, a 10,000 pixel submillimetre camera on the JCMT with eight sub-arrays.  SCUBA-2 operates at 450~$\micron$ and 850~$\micron$, like SCUBA, and can be susceptible to significant molecular line contamination.  %but could be more susceptible to molecular line contamination because of its ability to generate deeper maps than SCUBA.  
In regards to $^{12}$CO, both SCUBA-2 bandpass filters have a central transmission peak near a $^{12}$CO~line:  the 850~$\micron$ bandpass filter centre is at 347~GHz near the $^{12}$CO~3--2~line at 345.796 GHz and the 450~$\micron$ bandpass filter centre is 664~GHz near the $^{12}$CO~6--5~line at 691.473~GHz.  The proximity of the $^{12}$CO~line frequencies to the centres of the transmission peaks makes significant CO contamination in SCUBA-2 maps likely.

In this paper, we have calculated conversion factors used to convert maps of molecular line integrated intensity (K~km~s$^{-1}$) to maps of molecular line flux (mJy~beam$^{-1}$) that contaminates the dust continuum emission.  These conversion factors were calculated for $^{12}$CO~3--2 contributions to the 850~$\micron$ SCUBA-2 dust continuum emission and for $^{12}$CO~6--5 contributions to the 450~$\micron$ SCUBA-2 continuum.  Conversion factors were applied to HARP $^{12}$CO maps of NGC~1333, a region in the Perseus molecular cloud complex, and NGC~2071 and NGC~2024, regions in the Orion~B molecular cloud complex, to calculate the contamination directly by measuring fluxes and masses from a list of sources.  Once the $^{12}$CO contamination to the source fluxes was calculated, the sources with the highest contamination were analysed in more detail to determine the cause of the molecular flux contribution, e.g. molecular outflows or hot molecular gas from nearby stars.  

This paper has six sections.  In Section~\ref{method}, we give details of the calculation of the conversion factors considering different weather grades (Section \ref{factor}).  Section~\ref{results} presents the resulting conversion factors calculated for $^{12}$CO 3--2 (contributing to the 850~$\micron$ continuum band) and $^{12}$CO~6--5 (contributing to the 450~$\micron$ continuum band).  Section~\ref{flux} explains how the conversion factors were applied to HARP $^{12}$CO~3--2 maps to quantify the amount of contamination to the 850~$\micron$ SCUBA-2 dust continuum maps of three regions, NGC~1333, NGC~2071, and NGC~2024.  Section~\ref{fluxcalculations} and \ref{mass} introduce flux and mass calculations for a list of sources in the regions, and Section~\ref{outflows} presents further analysis of sources with high $^{12}$CO contamination.  Section~\ref{analysis} discusses the results, including the effects of environment and location (molecular outflows and hot ambient gas from nearby stars) on the $^{12}$CO source contamination.  This section also includes estimates of the $^{12}$CO~6--5 line to the 450~$\micron$ SCUBA-2 dust emission.  Lastly, Section~\ref{conclusions} summarises the conclusions drawn from this work.

\section{Method}
\label{method}

Molecular line emission is typically measured as an intensity or surface brightness in terms of the Rayleigh-Jeans (R-J) brightness temperature (in Kelvin), while the dust continuum fluxes are given in Janskys %($1 \ \mathrm{Jy} = 10^{26} \ \mathrm{W \ m}^2 \ \mathrm{Hz}^{-1} = 10^{23} \ \mathrm{erg \ s}^{-1} \ \mathrm{cm}^{-2} \ \mathrm{Hz}^{-1} $)
measured over the telescope beam area (Jy~beam$^{-1}$).  In order to convert $^{12}$CO line intensities to pseudo-continuum fluxes, the intensity of a molecular line must be converted into the flux of the line using the following relation \begin{equation}
F = \int I \ \mathrm{d}\Omega \approx I \Omega,
\label{eq:flux}
\end{equation} where $I$ is the intensity and $\Omega$ is the telescope beam area.  The intensity is measured as a main-beam brightness temperature $T_{\mathrm{MB}}$ in Kelvin and converted to intensity using

\begin{equation}
I_\nu = \frac{2\nu^2}{c^2} kT_{\mathrm{MB}} = \frac{2k}{\lambda^2} T_{\mathrm{MB}} 
\label{eq:I_nu},
\end{equation} where $\nu$ is the frequency, $\lambda$ is the wavelength, and $k$ is the Boltzmann constant.% ($\sim$~10$^{-23}$~kg~m$^{2}$~s$^{-2}$~K$^{-1}$).  

A narrow molecular line within a filter contributes flux over a smaller frequency range ($\nu_{\mathrm{ line}}$) than continuum emission
%in contrast to the continuum intensity that contributes 
across the filter.  To obtain the flux from the molecular line, the average intensity $\langle I \rangle$ must be calculated over the full filter band, i.e. \begin{equation}
\langle I \rangle = \frac{\int I_\nu\mathrm{(line)} \ g_\nu\mathrm{(line)} \ d\nu}{\int g_\nu \ d\nu},
\label{eq:TMB}
\end{equation} where $I_\nu \mathrm{(line)}$ is the intensity of the molecular line, $g_\nu \mathrm{(line)}$ is the filter passband (transmission) at the frequency of the molecular line, and $\int g_\nu \ d\nu$ is the integrated filter passband (transmission) across the full range of filter frequencies.  Using Equation \ref{eq:I_nu} and the  Doppler shift, $\Delta \nu / \nu = \Delta {\mathrm{v}} / c$, Equation \ref{eq:TMB} can be converted to $T_{MB}$:  \begin{equation}
\langle I \rangle = \frac{\frac{\nu}{c} \int I_\nu({\mathrm{v}}) \ g_\nu({\mathrm{v}}) \ d{\mathrm{v}}}{\int g_\nu \ d\nu} =  \frac{2k\nu^3}{c^3} \frac{g_\nu(\mathrm{line})}{\int g_\nu \ d\nu} \int T_{\mathrm{MB}} \ d{\mathrm{v}},
\end{equation} where $\int T_{\mathrm{MB}} \ d{\mathrm{v}}$ is the velocity integrated main-beam brightness temperature, or integrated intensity.  %(K~km~s$^{-1}$ or $10^{3}$~K~m~s$^{-1}$).  
Using these calculations of intensity, it follows from Equation~\ref{eq:flux} that \begin{equation}
\frac{F_\nu}{\mathrm{mJy \ beam}^{-1}} = \frac{2k\nu^3}{c^3} \frac{g_\nu \mathrm{(line)}}{\int g_\nu \ d\nu} \Omega_B \int T_{\mathrm{MB}} \ d{\mathrm{v}}
\label{eq:contribution}
\end{equation}  A similar calculation was used in \citet{2004MNRAS.349.1428S}.  

The main-beam brightness temperature $T_{\mathrm{MB}}$ is used rather than the antenna temperature $T_{\mathrm{A}}^\ast$ for analysing small-scale emission as long as the continuum calibration accounts for the same beam efficiencies as the molecular line emission.  This is true for the SCUBA-2 and HARP pairing at the JCMT.  The beam efficiencies are discussed further in Section~\ref{flux}.  %because the SCUBA-2 calibration accounts for the beam efficiencies in the flux conversion factors (FCFs) (discussed further in Section \ref{flux}). 
$T_{\mathrm{A}}^\ast$ is related to $T_{\mathrm{MB}}$ by the following:  \begin{equation}
T_A^\ast = \eta_{\mathrm{MB}} T_{\mathrm{MB}},
\end{equation}  where $\eta_{\mathrm{MB}}$ is the main-beam efficiency factor.  The efficiency factor that takes into account emission at larger scales is discussed in Section~\ref{beamarea}.  At $\sim$345 GHz with HARP on JCMT, $\eta_{\mathrm{MB}}$ is 0.61 \citep{2009MNRAS.399.1026B}.   %(JCMT Guide to Spectral Line Observing website)\footnote{http://docs.jach.hawaii.edu/JCMT/HET/GUIDE/het\_guide/het\_guide.html}.  
The telescope beam area, also discussed further in Section \ref{beamarea}, is measured in steradians (sterad) and obtained from the full-width-half-maximum (FWHM) $\theta_B$ of a Gaussian beam using $\Omega_B = 2 \pi \sigma^2$ where the FWHM $\theta_B = 2 \sqrt{2 \ln 2} \sigma$:  \begin{eqnarray}
\frac{\Omega_B }{\mathrm{sterad}} & =  & \frac{\pi}{4 \ln 2} \left(\frac{\theta_B}{''}\right)^2 \left(\frac{\pi}{180 \times 3600} \right)^2 
\label{eq:beam}
\end{eqnarray}

Using Equation \ref{eq:contribution}, a molecular line conversion factor, C, can be calculated to convert molecular line maps, measured in the velocity integrated main-beam temperature $\int {T_{\mathrm{MB}}} \ d\mathrm{v}$~(K~km~s$^{-1}$), into maps of line flux (mJy~beam$^{-1}$) that contributes to the observed continuum emission, %Therefore, conversion factors are calculated by dividing the flux $F_\nu$ by the velocity integrated main-beam temperature $\int {T_{\mathrm{MB}}} \ d\mathrm{v}$, yielding units of mJy~beam$^{-1}$~per~K~km~s$^{-1}$:
\begin{eqnarray}
\frac{C}{\mathrm{mJy \ beam}^{-1}{\mathrm{\ per \ K \ km \ s}^{-1}}} &=& \frac{F_\nu}{\int T_{MB} \ d{\mathrm{v}}} \\ \nonumber
&=& \frac{2k\nu^3}{c^3} \frac{g_\nu \mathrm{(line)}}{\int g_\nu \ d\nu} \Omega_B
\label{eq:conversion}
\end{eqnarray} where frequencies are measured in GHz and $1 \ \mathrm{Jy} = 10^{26} \ \mathrm{W \ m}^2 \ \mathrm{Hz}^{-1} = 10^{23} \ \mathrm{erg \ s}^{-1} \ \mathrm{cm}^{-2} \ \mathrm{Hz}^{-1}$.   Note that the beam size is wavelength dependent, where $\Omega_B \propto \lambda$.  The difference in beam size between the $^{12}$CO~line and the SCUBA-2 filter is not taken into account.

\subsection{Line Conversion Factors}
\label{factor}
\begin{figure*}
%\begin{minipage}{6in}
\centering
\includegraphics[width=5in]{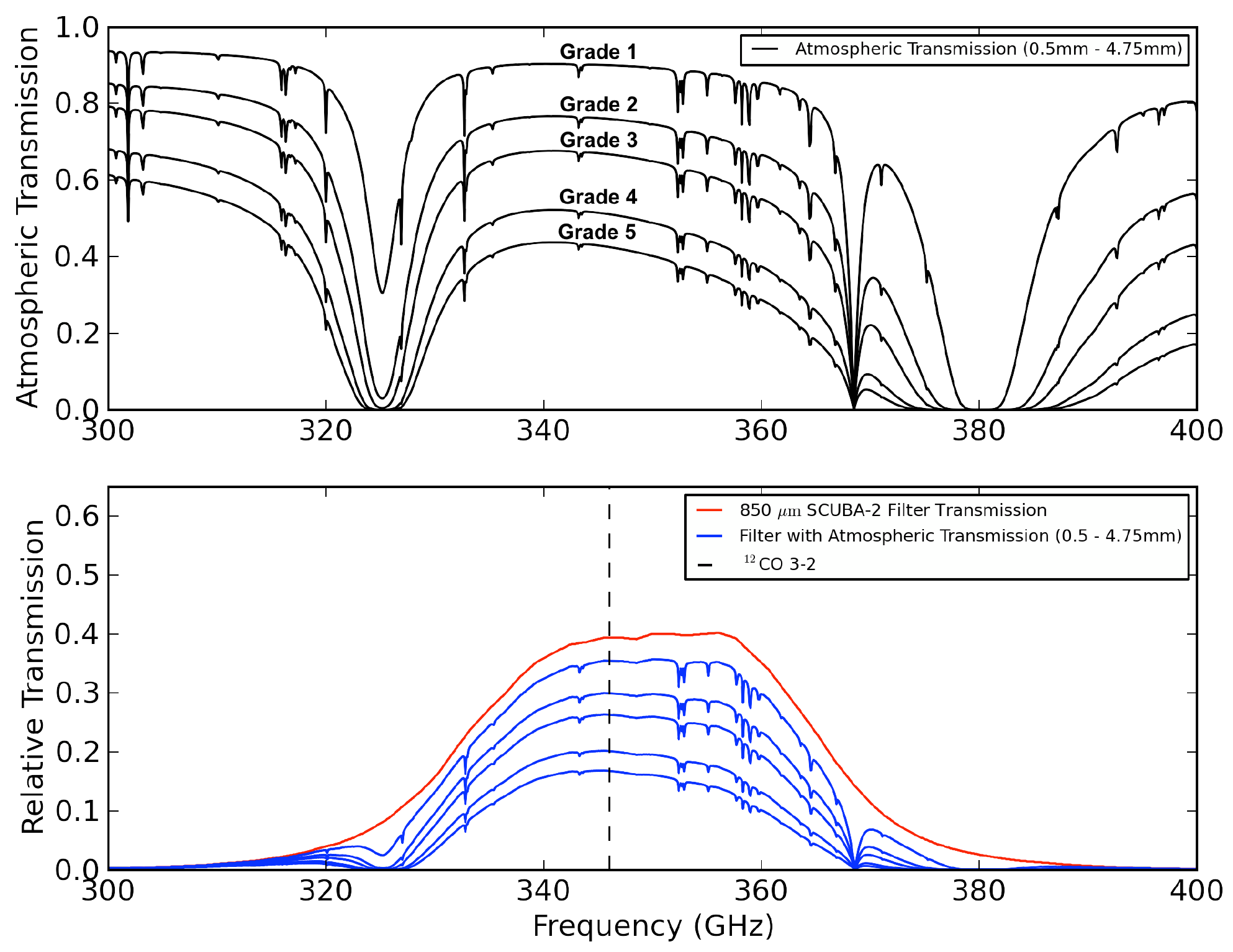}
\caption{{\em{Upper:}}  Plots of the atmospheric transmission at 300 to 400~GHz, given 0.5 to 4.75~mm of precipitable water vapour.  {\em{Lower:}}  The upper line is the profile of the SCUBA-2~850~$\micron$ filter, and the lines beneath represent the SCUBA-2 filter with the addition of the atmospheric transmission at varying water vapour levels.  The $^{12}$CO~3--2 line is plotted at 345.7960~GHz.  As shown in Table~\ref{table1}, the atmospheric transmission corresponds to different bands of weather used for observations.}
\label{fig:atmospheric}
%\end{minipage}
\end{figure*}

\begin{figure*}
%\begin{minipage}{6in}
\centering
\includegraphics[width=5in]{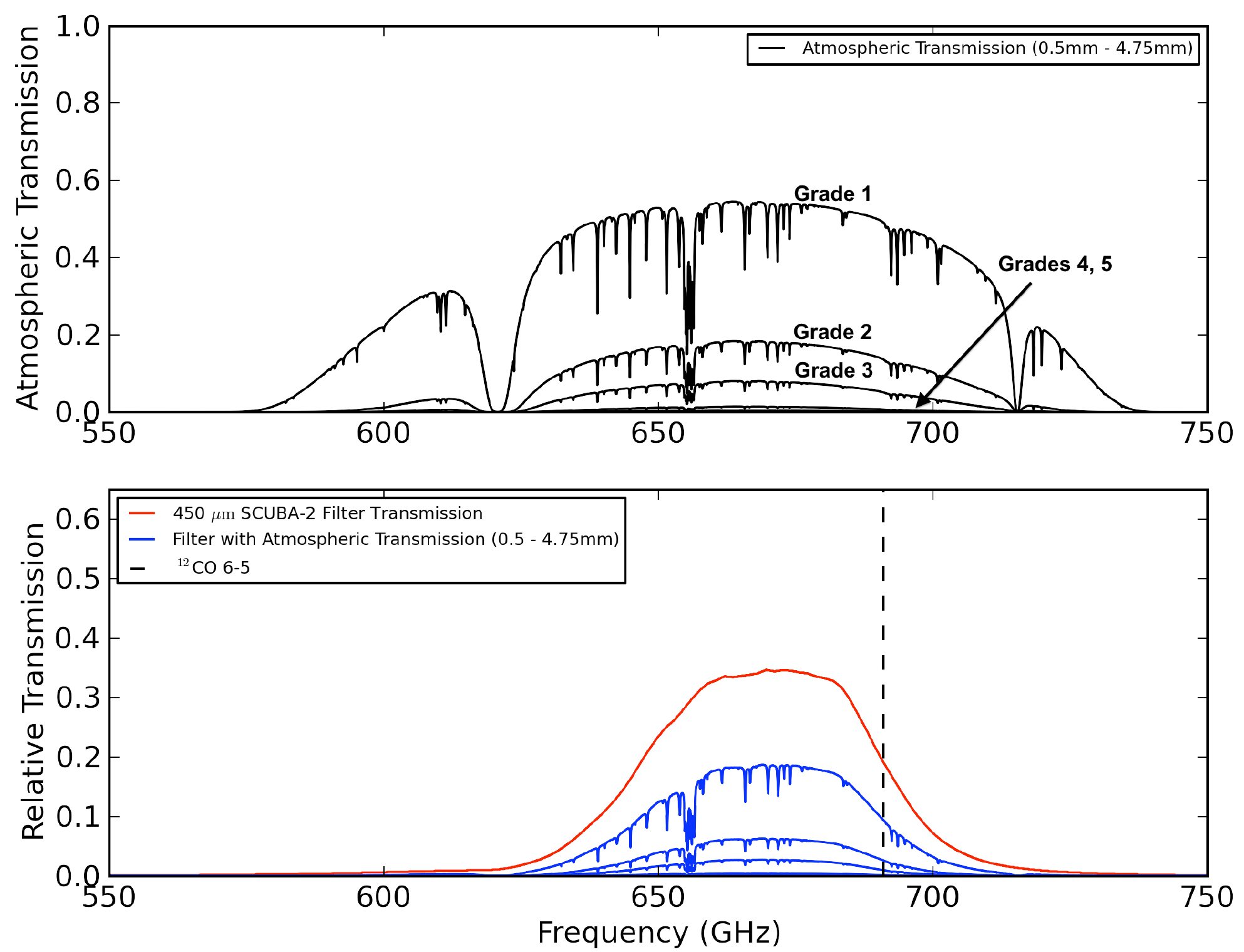}
\caption{{\em{Upper:}}  Plots of the atmospheric transmission at 550 to 750~GHz, given 0.5 to 4.75~mm of precipitable water vapour.  {\em{Lower:}}  The upper line is the profile of the SCUBA-2~450~$\micron$ filter, and the lower lines represent the SCUBA-2 filter with the addition of the atmospheric transmission at varying water vapour levels.  The $^{12}$CO~6--5 line is plotted at 691.4731~GHz.}
\label{fig:atmospheric450}
%\end{minipage}
\end{figure*}

To calculate the conversion factor C from Equation \ref{eq:conversion}, the SCUBA-2 filter profiles and the added atmospheric transmission were used to find ${\int g_\nu \ d\nu}$ and $g_\nu \mathrm{(line)}$.  The SCUBA-2 850 $\micron$ and 450~$\micron$ filter profiles are shown in the bottom plot of Figures \ref{fig:atmospheric} and \ref{fig:atmospheric450}.  The SCUBA-2 filter profiles are a result of stacking all of the filters (thermal and bandpass filters as well as the cryostat window and dichroic) that form the continuum bandpasses when combined with the atmosphere.  The bandpasses are the main filters defining the transmission window, where passbands are the range of frequencies with a signal passing through the filter and stopbands define frequency ranges with a signal attenuated by the filter.  The main infrared (IR; thermal) blocking filters are designed to block transmission at higher frequencies (IR and optical).  For this study, a constant value for these filters has been assumed due to the high transmission in our frequency range.  For further information, see the JCMT website regarding the cryostat window, filter and dichroic specification and measurements\footnote{\url{http://www.jach.hawaii.edu/JCMT/continuum/scuba2/filter/}}.  

The JCMT has a system that describes the atmospheric conditions ranging from weather grades 1-5.  The atmospheric conditions are based on precipitable water vapour (PWV) levels (in mm) that correspond to different sky opacities at 225 GHz, or $\tau_{225}$.  The relation between PWV and $\tau_{225}$ is the following (JCMT Telescope Overview website)\footnote{\url{www.jach.hawaii.edu/JCMT/overview/tel_overview}}:

\begin{equation}
\tau_{225} \approx 0.01 + (0.04 \times \mathrm{PWV})
\end{equation}  The JCMT weather grades are defined as:

\begin{itemize}
	\item {\bf{Grade 1:}}  PWV: $< 1$ mm, $\tau_{225}$: $< 0.05$
	
	\item {\bf{Grade 2:}}  PWV: 1 to 1.75 mm, $\tau_{225}$: 0.05 to 0.08	
		
	\item {\bf{Grade 3:}}  PWV: 1.75 to 2.75 mm, $\tau_{225}$: 0.08 to 0.12
	
	\item {\bf{Grade 4:}}  PWV: 2.75 to 4.75 mm, $\tau_{225}$: 0.12 to 0.20 	
	
	\item {\bf{Grade 5:}}  PWV: $> 4.75$ mm, $\tau_{225}$: $> 0.20$	
\end {itemize}

The continuum bandpass transmission of the filter profile and atmosphere combined varies depending on atmospheric conditions.  Therefore, the CO contamination was calculated based on these five weather grades.  Plots of atmospheric transmission corresponding to these conditions can be found in the top half of Figures~\ref{fig:atmospheric} and \ref{fig:atmospheric450} and are labelled according to the respective water vapour levels (for more detail, see CSO Atmospheric Transmission Interactive Plotter website).\footnote{\url{http://www.submm.caltech.edu/cso/weather/atplot.shtml}}  The SCUBA-2 850~$\micron$ and 450~$\micron$ filter profiles were multiplied by each individual atmospheric transmission profile to produce continuum bandpass profiles at each weather grade, shown in the bottom of Figures~\ref{fig:atmospheric} and \ref{fig:atmospheric450}.  

In Equation~\ref{eq:contribution}, $\int g_\nu \ d\nu$ is the integrated SCUBA-2 continuum bandpass and is calculated as the the sum of $g_\nu \times \delta \nu$ (where $\delta \nu$ is 0.01~GHz) at each corresponding frequency with units in GHz.  The transmission of $^{12}$CO, $g_\nu \mathrm{(line)}$, is the transmission of the SCUBA-2 850~$\micron$ continuum bandpass at 345.7960~GHz (the rest frequency of the $^{12}$CO~3--2~line) and the transmission of the SCUBA-2 450~$\micron$ continuum bandpass at 691.4731~GHz (the rest frequency of the $^{12}$CO~6--5~line).  To calculate conversion factors for redshifted lines, the frequency $\nu$ and transmission $g_\nu$(line) (from Figures~\ref{fig:atmospheric} \& \ref{fig:atmospheric450}) in Equation~\ref{eq:conversion} must be changed appropriately.  The SCUBA-2 beam size is calculated using Equation~\ref{eq:beam} assuming the main-beam FWHM~$\theta_B$ is 13.8$''$ at 850~$\micron$ and 8.3$''$ at 450~$\micron$.  % (JCMT SCUBA-2 Integration Times website)\footnote{\url{www.jach.hawaii.edu/JCMT/continuum/scuba2_integrat_time_calc.html}}.  
The possibility of an associated secondary beam is discussed in Section~\ref{beamarea}.

%The SCUBA-2 beam size in Equation~\ref{eq:beam} is calculated from the main-beam FWHM

\begin{table*}
%\begin{minipage}{8in}
\begin{tabular}{c c c c c c c c c c}
Band & Filter with & \bf{$\int g_\nu d\nu$} & \bf{$\theta_B$} & Line & $\nu$ & \bf{$g_\nu (\mathrm{line})$} & $\tau_{225}$& Weather & C\\
 & Atmospheric Trans. & GHz & $''$ & & GHz & & & Grade & mJy~beam$^{-1}$~per \\ 
 & (PWV in mm) & & & & & & & & K~km~s$^{-1}$ \\ \hline
 SCUBA-2 850 $\micron$ & 0.5 & 19.79 & 13.8 & $^{12}$CO 3--2 & 345.7960 & 0.58 & 0.03 & 1 & 0.63 \\ 
 & 1.5 & 15.63 & 13.8 & $^{12}$CO 3--2 & 345.7960 & 0.49 & 0.07 & 2 & 0.68 \\ 
 & 2.25  & 13.24 & 13.8 & $^{12}$CO 3--2 & 345.7960 & 0.43 & 0.10 & 3 & 0.70 \\ 
 & 3.75 & 9.56 & 13.8 & $^{12}$CO 3--2 & 345.7960 & 0.33 & 0.16 & 4 & 0.74 \\ 
 & 4.75 & 7.71 & 13.8 & $^{12}$CO 3--2 & 345.7960 & 0.28 & 0.20 & 5 & 0.77\\ 
&&&&&&&&&\\
SCUBA-2 450 $\micron$ & 0.5 & 8.49 & 8.3 & $^{12}$CO 6--5 & 691.4731 & 0.09 & 0.03 & 1 & 0.64 \\ 
 & 1.5 & 2.64 & 8.3 & $^{12}$CO 6--5 & 691.4731 & 0.02 & 0.07 & 2 & 0.57 \\ 
 & 2.25  & 1.10 & 8.3 & $^{12}$CO 6--5 & 691.4731 & 0.01 & 0.10 & 3 & 0.51 \\ 
 & 3.75 & 0.19 & 8.3 & $^{12}$CO 6--5 & 691.4731 & $< 0.01$ & 0.16 & 4 & 0.41 \\ 
 & 4.75 & 0.06 & 8.3 & $^{12}$CO 6--5 & 691.4731 & $< 0.01$ & 0.20 & 5 & 0.35 \\ \hline
\end{tabular}
\caption{Line contribution factors for $^{12}$CO lines in the SCUBA-2 850~$\micron$ \& 450~$\micron$ continuum bands.  }
\label{table1}
%\end{minipage}
\end{table*}	 

\subsection{Telescope Beam Area}
\label{beamarea}

%According to past research involving SCUBA \citep{2008ApJS..175..277D}, 
The beam profile of the original SCUBA instrument diverged from a single Gaussian and displayed a beam profile of two combined Gaussians:  a primary beam roughly corresponding to the assumed FWHM and a secondary beam of 40$''$ FWHM \citep{2008ApJS..175..277D}.  For the 450~$\micron$ maps, the primary beam had a 8.5$''$ FWHM with a 0.90 relative amplitude and the secondary beam had a 0.10 relative amplitude.  For the 850~$\micron$ maps, the primary beam had a 13.5$''$ FWHM with a 0.96 relative amplitude and the secondary beam had a 0.04 relative amplitude.

It is possible that the total beam of SCUBA-2 also includes a broader secondary component.  As explained in Section~\ref{factor}, the telescope beam areas for this study were calculated assuming FWHMs for the 450~$\micron$ (8.3$''$) \& 850~$\micron$ (13.8$''$) SCUBA-2 beams.  The primary beam is appropriate for studying the CO contamination in compact sources (small-scale emission), e.g.\ protostars and small outflows.  However, if the CO emission is both bright and extended, then it may be necessary to include the secondary beam in the calculation.  By fitting a two component Gaussian to coadded SCUBA-2 maps of Uranus, the 450~$\micron$ primary FWHM  is 8.7$''$ (relative amplitude 0.83) and secondary FWHM is 20.4$''$ (relative amplitude 0.17) and the 850~$\micron$ primary FWHM is 13.9$''$ (relative amplitude 0.97) and secondary beam FWHM is 39.1$''$ (relative amplitude 0.03).\footnote{Measured primary beam sizes are slightly larger than the sizes quoted in Section~\ref{factor}.  This is potentially due to small pointing shifts between coadded maps.}  The effective FWHM becomes 11.6$''$ and 15.3$''$ for 450~$\micron$ and 850~$\micron$ beams respectively.  This would cause the total beam area for 450~$\micron$ to be higher by a factor of 2.0 and the 850~$\micron$ total beam area to be higher by a factor or 1.2. 

If it is necessary to incorporate the secondary beam into the beam area calculation, then a new conversion factor can be calculated using Equation~\ref{eq:conversion}.  The conversion factors are directly proportional to the telescope beam area.  Assuming $C'$ is the the conversion factor with the inclusion of both a primary and secondary beam ($\Omega_B '$) and $C$ is the relation shown in Equation~\ref{eq:conversion}, then it follows from Equation~\ref{eq:conversion}:

\begin{equation}
C' = C \ \frac{\Omega_B '}{\Omega_B} = C \ \left( \frac{\mathrm{FWHM}'}{\mathrm{FWHM}} \right)^2 ,
\label{eq:newcontribution}
\end{equation} where %the beam area $ \Omega_B  \propto$ FWHM$^2$ and 
FWHM$'$ is the effective FWHM.  For large-scale and extended emission on scales significantly larger than the size of the telescope beam (greater than 13.8$''$, the 850~$\micron$~FWHM), it is also more appropriate to use $\eta_{\mathrm{fss}}$, the forward spillover and scattering efficiency, to calibrate the CO emission instead of the main-beam efficiency $\eta_{\mathrm{MB}}$.  The forward spillover and scattering efficiency measures the amount of coupling to an extended source up to a 30$'$ diameter (measured by observing the Moon).  Equation \ref{eq:newcontribution} becomes

\begin{equation}
C' = C \ \left( \frac{\mathrm{FWHM}'}{\mathrm{FWHM}} \right)^2 \frac{\eta_{\mathrm{MB}}}{\eta_{\mathrm{fss}}}
\label{eq:beamcontribution}
\end{equation}  The increase of telescope beam area caused from the inclusion of the secondary beam is somewhat counterbalanced by the use of $T_\mathrm{R} ^\ast = T_\mathrm{A} ^\ast / \eta_{\mathrm{fss}}$ rather than $T_{\mathrm{MB}}$, accounting for the more efficient telescope coupling to large-scale emission ($\eta_{\mathrm{fss}} = 0.71$ compared to $\eta_{\mathrm{MB}} = 0.61$).

%$\eta_{\mathrm{fss}}$, which has a higher value of efficiency (0.71) than $\eta_{\mathrm{MB}}$ (0.61) \citep{2009MNRAS.399.1026B}.  %\footnote{Taken from http://docs.jach.hawaii.edu/JCMT/HET/GUIDE/het\_guide/het\_guide.html}.  

%\begin{equation}
%C' = C \ \left( \frac{\mathrm{FWHM}'}{\mathrm{FWHM}} \right)^2
%\end{equation}q
The secondary beam and potential changes in conversion factors for large-scale emission are further discussed in Section~\ref{analysis}.

%If SCUBA-2 does have a secondary beam like SCUBA, then a secondary beam of 40$''$ would cause the beam size to increase by a factor of 1.3 for 850 $\micron$ and 3.7 for 450 $\micron$ since the conversion factors are directly proportional to the beam size.  In this study, we continued to use the expected values for the primary beam only because the SCUBA-2 beam size has not yet been finalised.

\section{Results}
\label{results}

The $^{12}$CO conversion factors, C, for SCUBA-2 are listed in Table~\ref{table1}.  The $^{12}$CO~3--2 conversion factors (in mJy~beam$^{-1}$~per~K~km~s$^{-1}$) range from 0.63 (Grade~1) to 0.77 (Grade~5) with a mid value of 0.70 (Grade~3).  The conversion factors change depending on the atmospheric conditions that affect the continuum bandpass profile.  Since each increase in $\tau_{225}$ causes the 850~$\micron$ continuum bandpass profile to become narrower with less overall transmission, $\int g_\nu d\nu$ in Equation~\ref{eq:contribution} shrinks faster than the transmission of $^{12}$CO, $g_\nu \mathrm{(line)}$.  Therefore, the $^{12}$CO~3--2 line contribution to the flux is lowest in Grade~1 weather and steadily increases with each step to Grade~5 weather.  %The $^{12}$CO~3--2 conversion factors (in mJy~beam$^{-1}$~per~K~km~s$^{-1}$) range from 0.70 (Grade~1) to 0.85 (Grade~5) with a mid value of 0.78 (Grade~3).  

For the 450~$\micron$ continuum bandpass profile, 
%the transmission rapidly decreases with each increase in $\tau_{225}$ and nears zero transmission in Grade 5 weather
the opposite trend is seen.  The $^{12}$CO line contribution to the 450~$\micron$ flux is highest in Grade~1 but steadily decreases with each step to Grade~5 weather.  In most cases, observations using SCUBA-2~450~$\micron$ would only be taken in Grade~1 to 3 weather due to the decreased transmission in higher weather grades.  The $^{12}$CO~6--5 conversion factors range from 0.64 (Grade~1) to 0.35 (Grade~5) with a mid value of 0.51 (Grade~3).  

Note that the contamination is expected to have different behaviour between the 450~$\micron$ and 850~$\micron$ filters.  The $^{12}$CO~3--2 line is in the centre of the 850~$\micron$ filter with the bulk of the transmission while the $^{12}$CO~6--5 line is close to the edge of the 450~$\micron$ filter with lower transmission.  The molecular line contribution to the 450~$\micron$ band decreases with weather grade because of the increasing attenuation of the $^{12}$CO~6--5 line.  

\section{Applications to Observations}
\label{flux}

The conversion factors calculated in Section~\ref{results} were applied to HARP $^{12}$CO~3--2 maps and compared to SCUBA-2~850~$\micron$ dust emission maps to measure the $^{12}$CO contamination directly.  Three different regions were used for this study:  NGC~1333, NGC~2071, and NGC~2024.  By quantifying the percentage of contamination to the dust continuum flux, we can determine regions more likely to be contaminated by CO (i.e.\ regions with molecular outflows or nearby stars). 

\subsection{Flux Calculations}
\label{fluxcalculations}

The continuum observations were taken with SCUBA-2 at 450~$\micron$ and 850~$\micron$ in each region in 2010 during the SCUBA-2 Shared Risk Observing (S2SRO) campaign when SCUBA-2 had two science grade arrays (one at 450~$\micron$ and one at 850~$\micron$) installed.  Observations were taken in Grade 2 weather conditions.% (Nutter et al. {\emph{in prep.}} \& Richer et al. {\emph{in prep.}}).  %SCUBA-2 is a 10,000 pixel submillimetre camera on the JCMT at Mauna Kea Observatory in Hawaii.  %The telescope has a 14.5 arcsec resolution at 850 $\mu$m and a 7.5 arcsec resolution at 450 $\mu$m. 

The Heterodyne Array Receiver Programme (HARP) is a 16 pixel array receiver (16 receptors separated by 30$''$ and a footprint of 2$'$).  The $^{12}$CO~3--2 data cube for NGC~1333 was observed in January 2007 using raster mapping techniques (for details see \citealt{2010MNRAS.408.1516C}).
%  The data was taken in good observing conditions ($\tau_{225} = 0.125$ or Grade 2/3 weather).  The Auto-Correlation Spectral Imagining System (ACSIS), which is the background correlator, was set to supply 1000 MHz of bandwidth that was split into 2048 spectral channels with a width of 0.42 km s$^{-1}$ (488 kHz).  
The $^{12}$CO 3--2 data cubes for NGC 2071 and NGC 2024 were observed in November 2007 using raster mapping techniques as well (see \citealt{2010MNRAS.401..204B}).  
%The data was taken in good observing conditions ($\tau_{225} = 0.075$ or Grade 2 for NGC2071 and $\tau_{225} = 0.08$-$0.09$ or Grade2/3 for NGC2024).  ACSIS was set to supply 250 MHz bandwidth that was split into 4096 spectral channels with a width of 0.05 km s$^{-1}$ (61 kHz).  The velocity channels 
Both datacubes were rebinned to 0.42 km s$^{-1}$ velocity channels and converted to $T_{\mathrm{MB}}$ using a main-beam efficiency $\eta_{\mathrm{MB}}$ of 0.61. % of NGC1333 when analysing the $^{12}$CO spectra (see Section \ref{outflows}). 

Similar to other ground-based bolometer arrays, the limited, single-subarray version of SCUBA-2 available for S2SRO reproduced maps that are not sensitive to large-scale emission, in this case on scales larger than the single subarray field-of-view ($\sim$4$'$).  However, the HARP $^{12}$CO maps still contain this large-scale flux.  A simple application of the CO conversion factors from Table~\ref{table1} to the HARP maps would retain the large-scale structure and overestimate the CO contamination.  In order to account for the spatial filtering inherent in bolometer array reconstruction in a simple way and subtract the large-scale flux from the maps, a Gaussian smoothing mask (GSM) filter was applied to both the $^{12}$CO HARP integrated intensity maps (in K~km~s$^{-1}$) and the 850~$\micron$ maps (mJy~beam$^{-1}$) of the regions.  The GSM filter was designed to minimise emission from structure on scales inaccessible to SCUBA-2 at this time.  For the HARP maps, GSM filters were created by convolving a HARP $^{12}$CO contamination map directly with a Gaussian a few arc minutes FWHM in size and subtracting the resulting smoothed map from the original map.  For the SCUBA-2 maps, it was necessary to first create a thresholded map for masking bright protostars and convolve the thresholded map with a Gaussian the same FWHM in size.  The resulting smoothed map was then subtracted from the original map.  

\begin{figure*}
%\begin{minipage}{7in}
\centering

\includegraphics[width=3in]{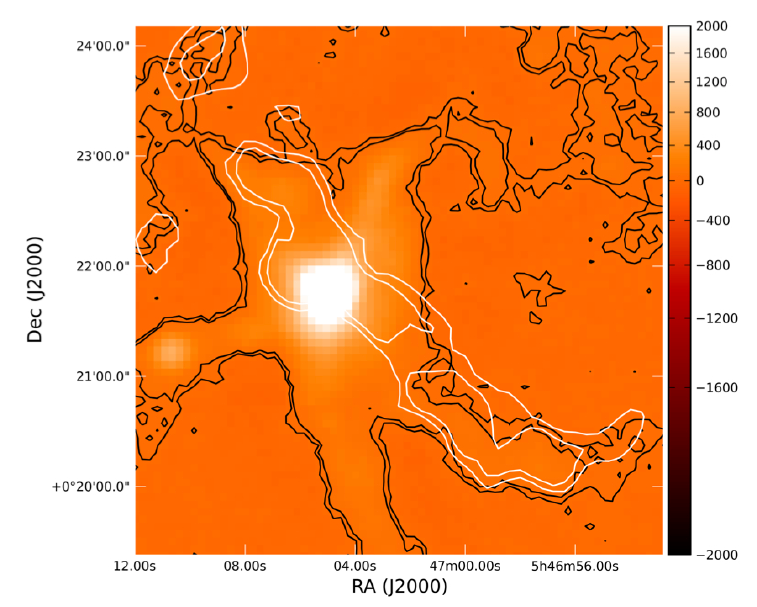}
\includegraphics[width=3in]{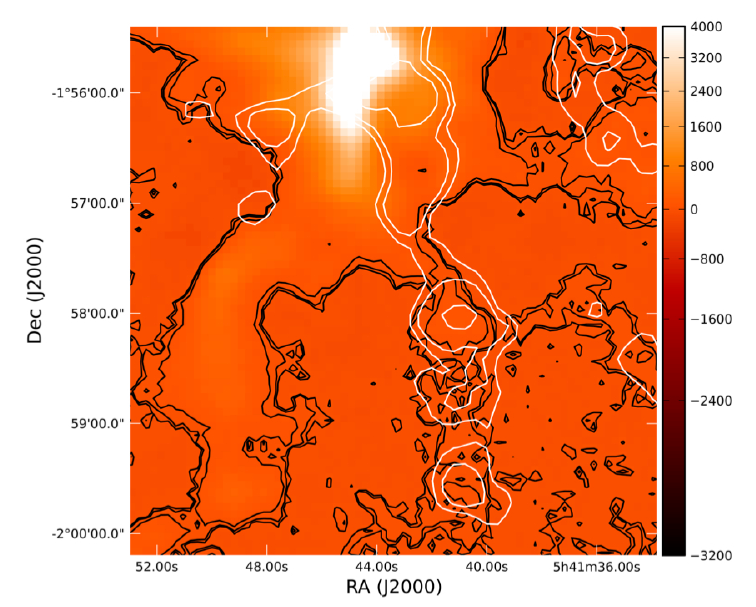}
\includegraphics[width=3in]{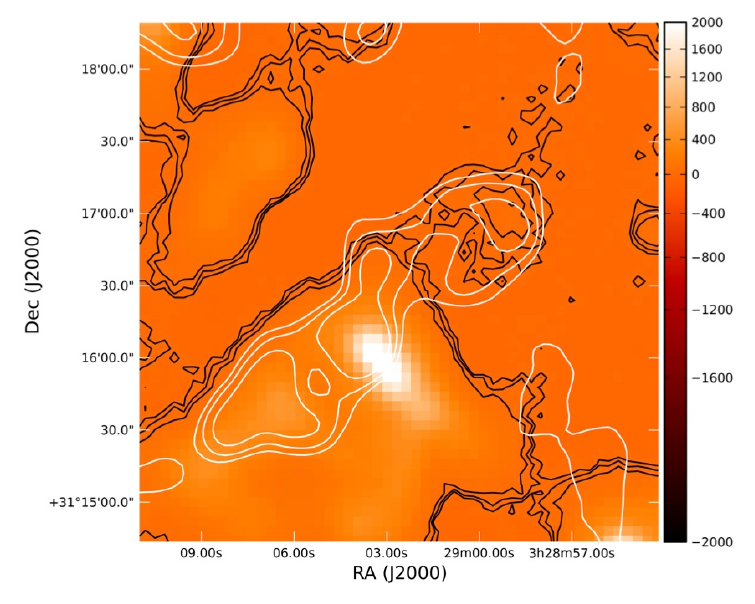}

\caption{Regions in the SCUBA-2 850~$\micron$ GSM-filtered maps of NGC~2071, NGC~2024, and NGC~1333 where $^{12}$CO emission contributes strongly to the 850~$\micron$ flux.  Black contours correspond to the SCUBA-2 850~$\micron$ dust continuum maps and white contours correspond to HARP~$^{12}$CO~3--2 contamination maps.  %Contour levels at 10, 20, 50, 100, \& 200~mJy~beam$^{-1}$.  
{\em{Top Left:}}  Close-up of LBS-MM18 (NGC2071-IRS) and corresponding outflow.  Both sets of contours correspond to flux at 20 and 45 mJy~beam$^{-1}$.  Noticeable $^{12}$CO flux contribution in the lower right lobe of the molecular outflow. {\em{Top Right:}}  Close-up of FIR~1-7 and corresponding outflow.  Both contours correspond to flux at 10, 20, and 45 mJy~beam$^{-1}$.  Noticeable $^{12}$CO flux contribution in the lower lobe of the molecular outflow. {\em{Bottom:}}  Close-up of  SVS13 and corresponding outflow.  Both sets of contours correspond to flux at 10, 20, and 35 mJy~beam$^{-1}$.  Noticeable $^{12}$CO flux contribution in the right lobe of the molecular outflow. }
%\caption{The SCUBA-2 850~$\micron$ GSM-filtered map of NGC~1333 region with contours from the HARP $^{12}$CO 3--2 contamination map of NGC~1333 overplotted.  The colour bar is shown in units of flux (mJy~beam$^{-1}$).  The black contours denote $^{12}$CO~3--2 emission levels at 20, 30, 40, 50, 70, 90, 120, 150, \& 200 mJy~beam$^{-1}$.}
\label{fig:codust_ngc1333}
%\end{minipage}
\end{figure*}

A $1'$ FWHM Gaussian was chosen for generating GSM maps after analysing $1'$ to $3'$ GSM filter sizes, further discussed in Section~\ref{GSM}.  Figure~\ref{fig:codust_ngc1333} shows examples of the SCUBA-2 GSM processed maps for protostellar cores LBS-MM18 (NGC~2071-IRS) in NGC~2071 (see \citealt{2001A&A...372L..41M}), FIR~1-7 in NGC~2024(see \citealt{1989MNRAS.241..231R, 2010MNRAS.401..204B}), and SVS13 in NGC~1333 (see \citealt{2007A&A...468.1009H}).  The brightest CO features in the maps are the result of molecular outflows driven by the dense cores in the regions.  Dust emission contours in the outflows of these regions clearly follow the $^{12}$CO~3--2 emission, indicating the CO contamination is strong enough to be directly detected in the dust continuum.  

To study the CO contamination quantitatively, aperture photometry with a 15$''$ radius was applied to lists of known submillimetre sources, listed in Table~\ref{table:source_list} and further discussed in Sections~\ref{ngc1333_flux}, \ref{ngc2071_flux}, and \ref{ngc2024_flux}.  A 15$''$ aperture radius was chosen based on source proximity and the possibility of the aperture diameter extending to a neighbouring source.  Integrated flux densities are calculated by assuming a sky background of zero with flux uncertainties based on the sky RMS and include a correction for the Gaussian beam \citep{2006ApJ...638..293E}.  Therefore, a point-source has the same integrated flux density in any size aperture.  

\subsubsection{Application to NGC~1333}
\label{ngc1333_flux}

NGC~1333 is a reflection nebula in the Perseus molecular cloud and is characterised by early stage star formation of age less than 1~Myr \citep{1996AJ....111.1964L, 2004AJ....127.1131W}.  The flux calibration for the S2SRO maps of this region was the CRL618 nebula and pointing checks were from the active galactic nucleus 3C84.  A flux conversion factor (FCF) of 500~Jy~beam$^{-1}$~pW$^{-1}$ was used for NGC~1333 to convert the maps into mJy~beam$^{-1}$.  Sources were chosen from a list of cores in NGC~1333 \citep{2007A&A...468.1009H} that had been previously identified in the submillimetre using SCUBA \citep{2005A&A...440..151H} and Bolocam \citep{2006ApJ...638..293E}  with a total of 35 sources in the area covered by the SCUBA-2 map.  These sources include a mixture of protostellar and starless cores.  For further information regarding HARP observations, see \citet{2010MNRAS.408.1516C}.

\begin{figure*}
%\begin{minipage}{6in}
\centering
\includegraphics[width=5in]{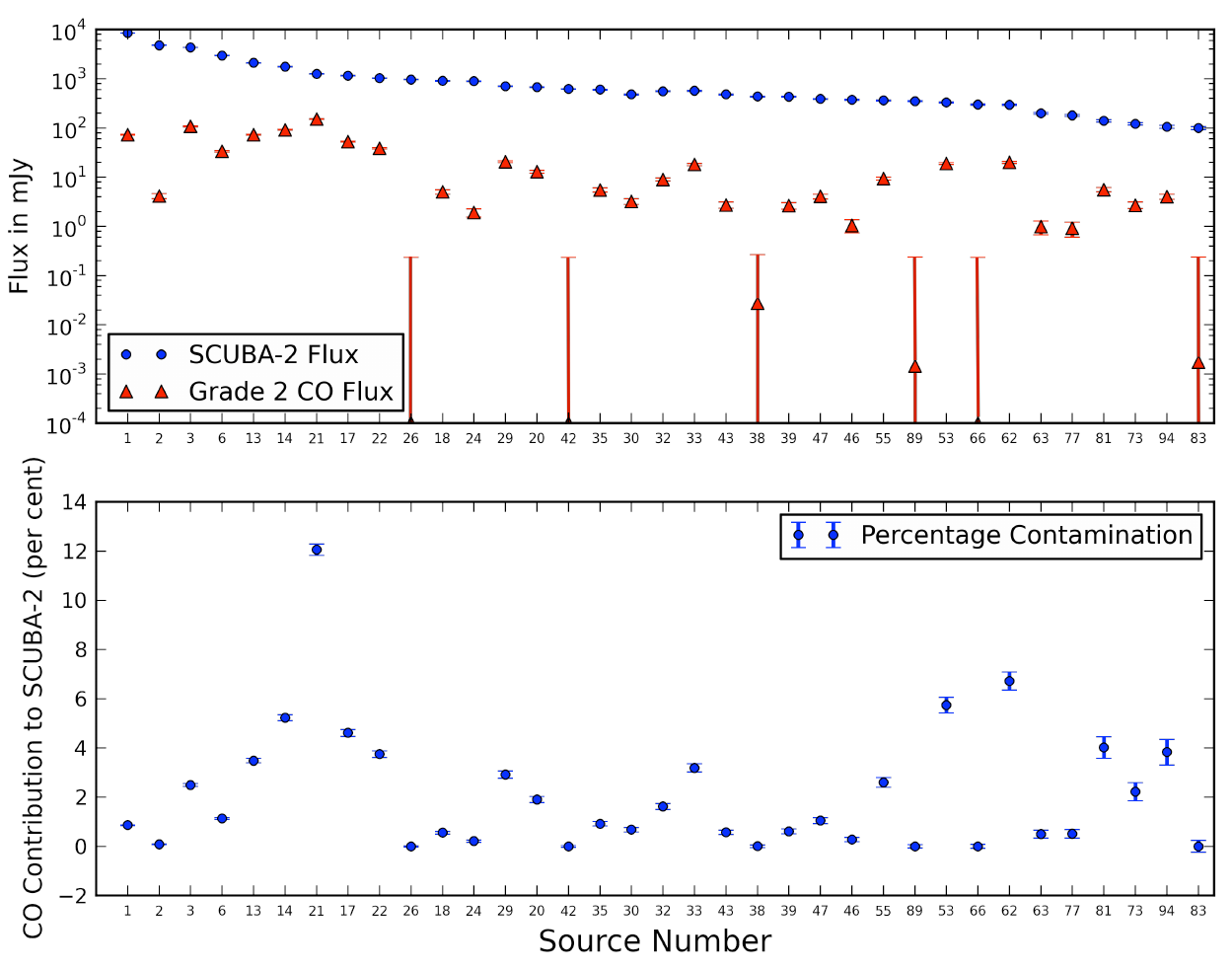}
\caption{{\em{Top:}}  The SCUBA-2 source fluxes calculated from the 850~$\micron$ continuum and $^{12}$CO~3--2 contamination maps (Grade~2 weather) of NGC~1333.  Note several sources (26, 42, and 66) have $^{12}$CO flux contributions of 0~mJy~beam$^{-1}$.  {\em{Bottom:}}  The percentage contribution to the SCUBA-2 fluxes from the $^{12}$CO contamination maps.  Numbers are given arbitrarily to the sources and were based on the original list of SCUBA and Bolocam cores \citep{2007A&A...468.1009H}.}
\label{fig:NGC1333_flux_compare}
%\end{minipage}
\end{figure*}

Figure~\ref{fig:NGC1333_flux_compare} shows the source fluxes from the SCUBA-2 850~$\micron$ and $^{12}$CO~3--2 Grade~2 contamination maps and the percentage contribution of $^{12}$CO~3--2 flux to 850~$\micron$ SCUBA-2 flux.  It should be noted that all of the sources have $^{12}$CO contributions less than 20~per~cent and every source except one (source~21) has a contribution less than 10~per~cent.  Source~21 is further discussed in Section~\ref{outflows}.  

\subsubsection{Application to NGC~2071}
\label{ngc2071_flux}

NGC~2071 is a region in the Orion~B molecular cloud.  Sources were chosen from a list of young stellar objects that had been previously identified using SCUBA \citep{2007MNRAS.374.1413N} with a total of 50 sources in the area covered by the SCUBA-2 map.  A flux conversion factor of 685~Jy~beam$^{-1}$~pW$^{-1}$ was used for NGC~2071 as well as NGC~2024 (FCF value valid for October 2010 reduction, equivalent to Nutter et al, {\em{in prep.}}).  For further information on HARP observations of NGC~2071, see \citet{2010MNRAS.401..204B}.

\begin{figure*}
%\begin{minipage}{6in}
\centering
\includegraphics[width=5in]{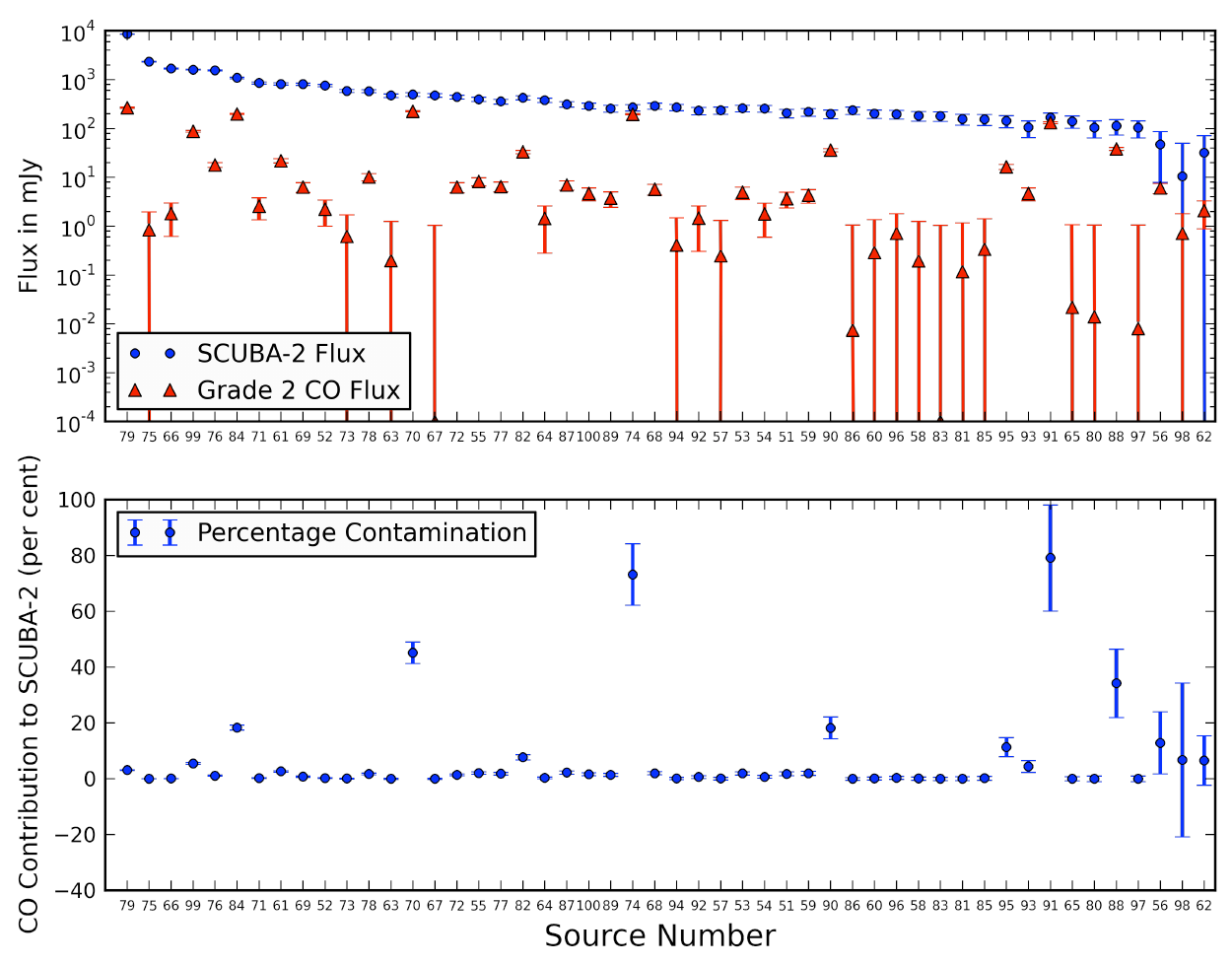}
\caption{{\em{Top:}}  The SCUBA-2 source fluxes calculated from the 850~$\micron$ continuum and $^{12}$CO~3--2 contamination maps (Grade~2 weather) of NGC~2071.  Note source 67 has a $^{12}$CO flux contribution of 0~mJy~beam$^{-1}$.  {\em{Bottom:}}  The percentage contribution to the SCUBA-2 fluxes from the $^{12}$CO contamination maps.  Numbers are given arbitrarily to the sources and were based on the original list of cores from \citet{2007MNRAS.374.1413N}.}
\label{fig:NGC2071_flux_compare}
%\end{minipage}
\end{figure*}

Figure~\ref{fig:NGC2071_flux_compare} shows the source fluxes from the SCUBA-2 850~$\micron$ and $^{12}$CO~3--2 Grade~2 contamination maps and the percentage contribution of $^{12}$CO~3--2 flux to SCUBA-2 850~$\micron$ flux.  Note that the majority of sources have $^{12}$CO contributions of less than 20~per~cent in Grade 2 weather.  Four sources (sources~70, 74, 88, and 91) have $^{12}$CO contributions greater than 20~per~cent (ranging from 34 to 79~per~cent).  Sources with a higher $^{12}$CO contamination are further discussed in Section~\ref{outflows}. 

\subsubsection{Application to NGC~2024}
\label{ngc2024_flux}

NGC~2024 is another emission nebula in the Orion~B molecular cloud.  Sources were chosen from a list of young stellar objects in NGC~2024 that had been previously observed using SCUBA \citep{2007MNRAS.374.1413N} with a total of 24 sources in the area covered by the SCUBA-2 map.  As stated in Section \ref{ngc2071_flux}, a FCF of 685~Jy~beam$^{-1}$~pW$^{-1}$ was used to correspond with current studies of Orion B (FCF value valid for October 2010 reduction, equivalent to Nutter et al, {\em{in prep.}}).  For further information on HARP observations of NGC~2024, see \citet{2010MNRAS.401..204B}.

\begin{figure*}
%\begin{minipage}{6in}
\centering
\includegraphics[width=5in]{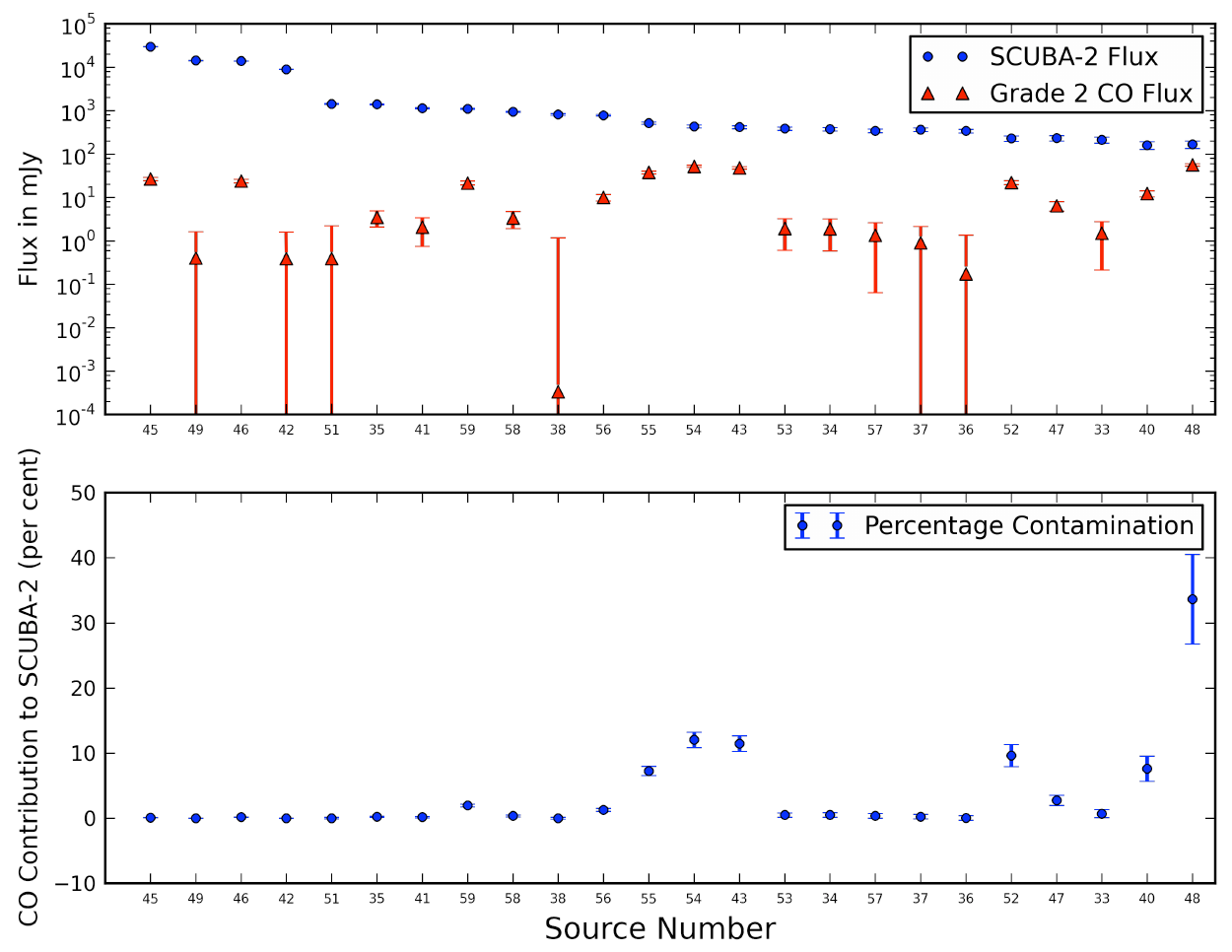}
\caption{{\em{Top:}}  The SCUBA-2 source fluxes calculated from the 850~$\micron$ continuum and $^{12}$CO 3--2 contamination maps (Grade~2 weather) of NGC~2024.  {\em{Bottom:}}  The percentage contribution to the SCUBA-2 fluxes from the $^{12}$CO contamination maps.  Numbers are given arbitrarily to the sources and were based on the original list of cores from \citet{2007MNRAS.374.1413N}.}
\label{fig:NGC2024_flux_compare}
%\end{minipage}
\end{figure*}

Figure~\ref{fig:NGC2024_flux_compare} shows the source fluxes from the SCUBA-2 850~$\micron$ and $^{12}$CO~3--2 Grade~2 contamination maps  and the percentage contribution of $^{12}$CO~3--2 flux to SCUBA-2 850~$\micron$ flux.  Note that the majority of sources have $^{12}$CO contributions of less than 20~per~cent.  One source (source~48) has a $^{12}$CO contribution more than 20~per~cent (34~per~cent contamination), further discussed in Section~\ref{outflows}.  

\subsubsection{GSM Analysis}
\label{GSM}
For SCUBA-2, GSM filters were created by applying upper thresholds to the original SCUBA-2 maps which acted as a mask for source emission and convolving the thresholded maps with a Gaussian a few arcminutes in FWHM size (similar to \citealt{2005ApJ...625..891R, 2006ApJ...646.1009K}); this is a standard technique for the SCUBA-2 data.  Negative regions of flux, known as negative `bowls,' surround very strong sources in the SCUBA-2 maps and are produced in the map reconstruction process (see \citealt{2000ApJ...545..327J}).  SCUBA-2 thresholding was necessary to prevent introducing new negative bowls in the image caused by smoothing and subtracting bright continuum sources.  Without thresholding, artificial negative bowling would have been further added to the map, causing negative flux to lower source fluxes and increase the calculated $^{12}$CO~3-2 contamination.  To test the effects of changing the filter size, S2SRO maps were analysed with an upper threshold of 15 mJy~beam$^{-1}$ for NGC~1333 and 30~mJy~beam$^{-1} $for NGC~2071 and NGC~2024 (used to mask out bright sources) and $1'$ to $3'$ FWHM Gaussian smoothing.  Aperture fluxes from each of the GSM maps ($1'$, $2'$, and $3'$) and the original, unsmoothed S2SRO maps were found to agree within uncertainties, indicating that the emission on scales which would have been affected by the GSM filter had been filtered out by the SCUBA-2 map reconstruction.  The $1'$ and $2'$ GSM filters were further analysed with application to the HARP $^{12}$CO maps based on the similarity between S2SRO and SCUBA maps, for which scales greater than $2'$ are known to be poorly reproduced \citep{2007A&A...468.1009H}.  %The GSM filter has the additional benefit of reducing the negative bowling effects around bright sources in the SCUBA-2 maps.  

For the HARP $^{12}$CO maps, thresholding was not required because %threshold maps were not used to generate GSM filters 
$^{12}$CO mainly traces molecular outflows which have bright, extended structures on scales not fully reproduced by SCUBA-2.  To generate the $^{12}$CO GSM maps, GSM filters with $1'$ and $2'$ FWHM Gaussians were directly applied and subtracted from the original maps.  Negative flux regions in the final HARP maps resulting from the oversubtraction of background flux estimated by the GSM filter was set to 0~mJy~beam$^{-1}$ to prevent biasing the source fluxes in the aperture photometry process.  The S2SRO $1'$ and $2'$ GSM maps were subtracted from the corresponding $^{12}$CO GSM maps (Grade~2 contamination) to analyse the GSM filter effectiveness in matching the spatial filtering of the SCUBA-2 maps.  %, shown in Figures~\ref{fig:GSM_ngc1333} to \ref{fig:GSM_ngc2024}.  
Positive flux in the residuals indicates higher $^{12}$CO flux than 850~$\micron$ dust continuum flux, suggesting that the GSM filter size should be scaled down to subtract smaller scale emission.  In each of the regions, the $2'$ GSM residuals were found to overestimate the $^{12}$CO flux contribution to the dust continuum.  On average, the $2'$~GSM map residuals were 1.4 to 1.5~times greater than the $1'$~GSM.  The $1'$ GSM filters were applied to both the S2SRO and HARP maps for consistency in eliminating flux on scales of $1'$ and above.
 
With the full complement of subarrays, SCUBA-2 is likely to recover more large-scale structure and continuum fluxes may increase further.  A comparison of $^{12}$CO contamination on scales of 1$'$ or greater will have to wait for full SCUBA-2 operations.

\subsection{Mass Calculations}
\label{mass}

%{\color{red}{(Why calculate masses?  A few sentences of motivation and aims and earlier references eg Hildebrand 1985 or Spitzer book.)}}

The calculation of the dust continuum flux from pre- and protostellar sources in a molecular cloud can be used to obtain source masses \citep{1983QJRAS..24..267H}.  Depending on the molecular cloud environment surrounding the sources, contamination from the $^{12}$CO line emission may affect low- and high-mass sources, leading to a varying level of source contamination.  Therefore, the masses of sources were calculated using the relation between the dust and gas mass and the total source dust continuum flux (e.g. \citealt{2004MNRAS.349.1428S, 2006ApJ...638..293E}),  \begin{equation}
M = \frac{S_{850}D^2}{\kappa_{850}B_{850}(T_{\rm {d}})},
\end{equation} where $S_{850}$ is the flux from 15$''$ radius aperture photometry at 850 $\micron$, $D$ is the distance to the source, $\kappa_{850}$ is the dust opacity at 850 $\micron$ and $B_{850}(T_{\mathrm{d}})$ is the Planck function at 850 $\micron$ for the dust temperature $T_{\mathrm{d}}$.

\subsubsection{Mass Calculations for NGC~1333}  

%{\color{red}{Does either your choice of Temp or dust opacity (kappa) effect the before or after masses?}}

For NGC1333, a distance of 250~pc was assumed for the mass calculations.  \citet{2007A&A...468.1009H} used a distance of 320~pc, which would increase masses by a factor of 1.6.  A temperature of 10~K was used as an estimate of the dust temperature, where dense regions that do not have internal heating are colder on the inside and warmer on the outside \citep{2001ApJ...557..193E}.  Cores with internal heating are warmer in the inner regions.  For example, Class~0 and Class~I protostars are found from models to have $\sim T_{\mathrm{d}} = 15$~K \citep{2002ApJ...575..337S, 2003ApJS..145..111Y}, but most of the dust mass is found in areas of lower temperatures.  To cover pre- and protostellar sources, $T_{\mathrm{d}} = 10$~K is a commonly used average.  It should be noted that this value can overestimate the masses of protostellar sources by a factor of 2 to 3 if the temperature is warmer \citep{2006ApJ...638..293E}. 

The dust opacity $\kappa_{850}$ is also uncertain for individual regions.  \citet{2007A&A...468.1009H} assumed a dust opacity of 0.012~cm$^2$g$^{-1}$ for the 850~$\micron$ SCUBA dust emission maps of NGC~1333 based on a gas/dust ratio of 161 (see \citealt{1994A&A...291..943O}).  
%by assuming a gas/dust ratio of 161 by mass.  
This dust opacity is at the low end of the assumed values and a dust opacity of 0.02~cm$^2$g$^{-1}$ at 850~$\micron$ could have been used \citep{2006ApJ...646.1009K}.  %\citet{2000MNRAS.315..115D} and \citet{2004MNRAS.349.1428S} assumed a dust opacity of 0.0077 cm$^2$g$^{-1}$ for galactic studies at 850 $\mu$m.  
Here, we choose a dust opacity of 0.012~cm$^2$g$^{-1}$.  If 0.02~cm$^2$g$^{-1}$ were used, then our masses would decrease by a factor of 1.7.

\begin{figure*}
%\begin{minipage}{6in}
\centering
\includegraphics[width=5in]{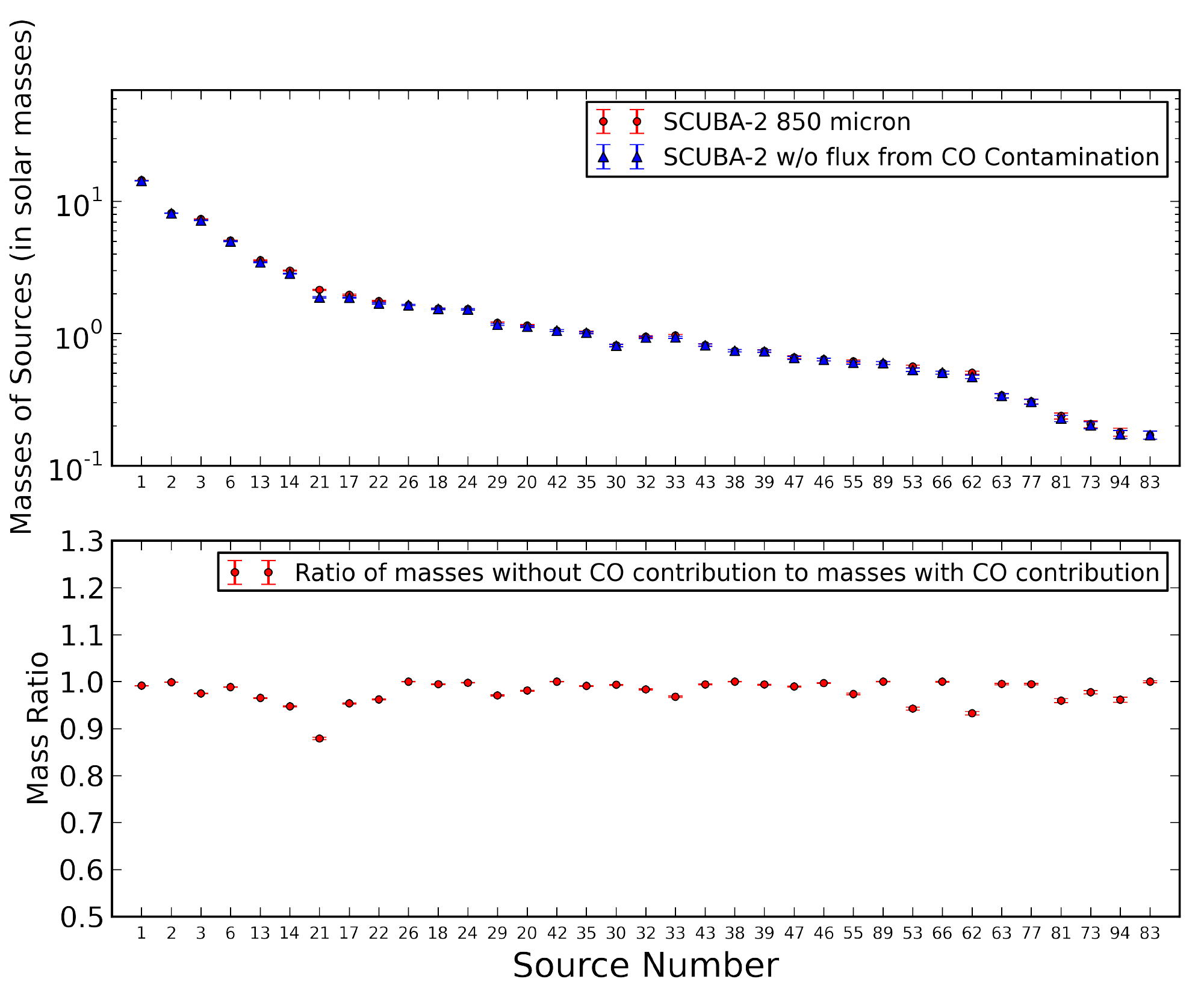}
\caption{{\em{Top:}}  Mass calculations (in solar masses) of the different sources in NGC~1333.  The masses were calculated using the SCUBA-2 850~$\micron$ map and then recalculated excluding the flux contribution from $^{12}$CO in different atmospheric conditions.  {\em{Bottom:}}  Ratio of the masses calculated from the flux without to with the $^{12}$CO contribution.  In both plots, uncertainties are calculated only from the source fluxes and do not include absolute calibration uncertainties.}% of $\pm$20~per~cent.}
\label{fig:NGC1333_mass_compare}
%\end{minipage}
\end{figure*}

Masses were calculated from source fluxes obtained from continuum emission with and without $^{12}$CO contamination taken into account.  The bottom portion of Figure~\ref{fig:NGC1333_mass_compare} shows the ratio between these masses.   
%calculated excluding $^{12}$CO contamination and the original masses calculated from SCUBA-2 fluxes (that include $^{12}$CO contamination).  
Due to the CO flux contamination, the calculated source masses are being overestimated by up to a factor of 1.2.  

\subsubsection{Mass Calculations for NGC 2071 and NGC 2024}

For NGC~2071 and NGC~2024, parameters from past mass estimates \citep{2007MNRAS.374.1413N} were used to calculate source masses.  A distance of 400~pc was assumed for both regions \citep{1994A&A...289..101B} and a temperature of 20~K was assumed as an estimate of the dust temperature \citep{1996A&A...312..569L, 2001ApJ...556..215M, 2006ApJ...653..383J}. An 850~$\micron$ dust opacity of 0.01~cm$^2$~g$^{-1}$  was used \citep{2003cdsf.conf..127A, 1996A&A...314..625A, 1999MNRAS.305..143W}, similar to the dust opacity used for NGC~1333.  As in NGC~1333, the assumption of a single temperature for each source does introduce a potential bias in the masses.  If 10~K was assumed, as for NGC~1333, then masses for NGC~2071 and NGC~2024 sources would be larger by a factor of 2. %up to 4.8 and 1.5 respectively.  

\begin{figure*}
%\begin{minipage}{6in}
\centering
\includegraphics[width=5in]{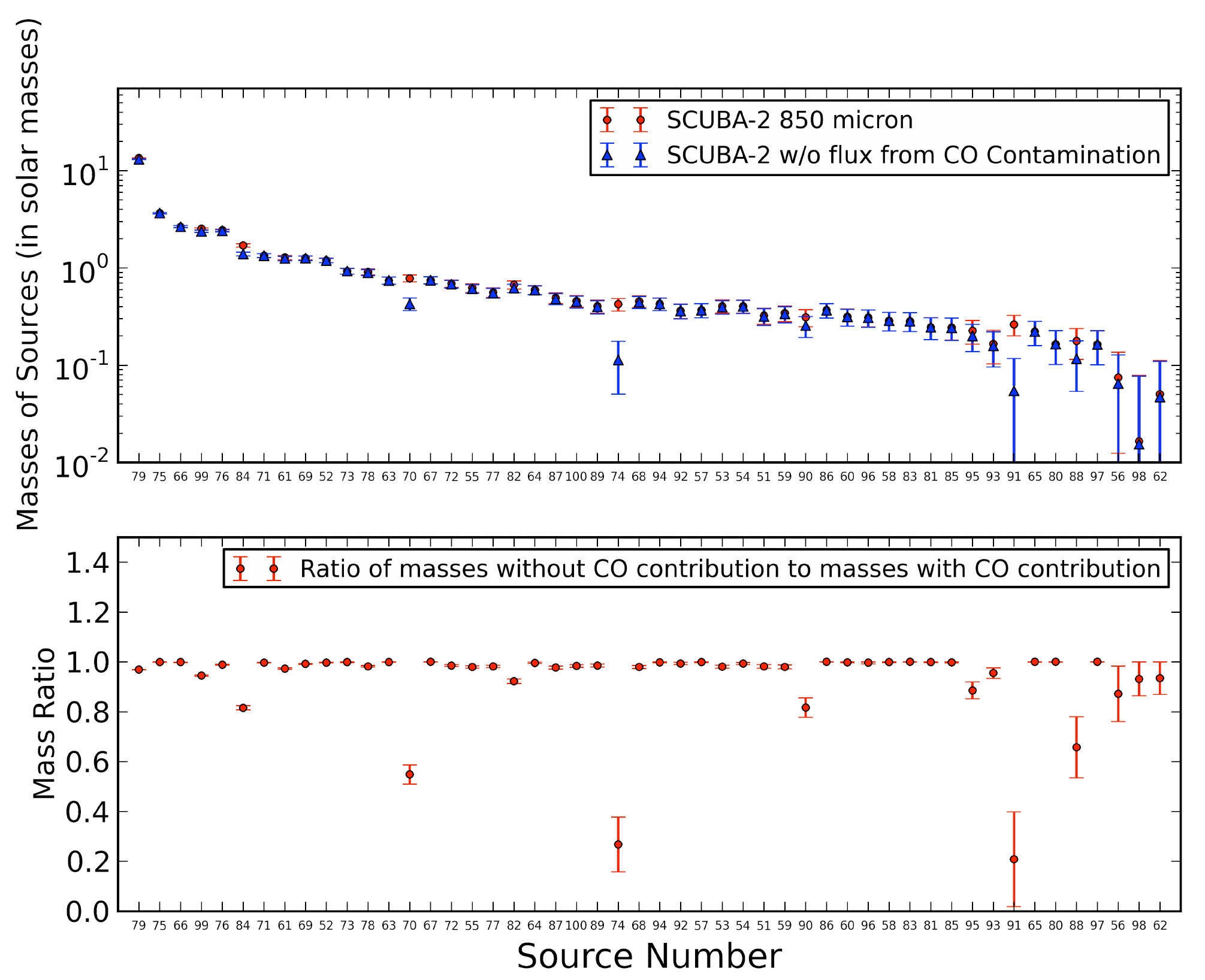}
\caption{{\em{Top:}}  Mass calculations (in solar masses) of the different sources in NGC~2071.  The masses were calculated using the SCUBA-2 850~$\micron$ map and then recalculated excluding the flux contribution from $^{12}$CO in different atmospheric conditions.  {\em{Bottom:}}  Ratio of the masses calculated from the flux without to with the $^{12}$CO contribution.  In both plots, uncertainties are calculated only from the source fluxes and do not include absolute calibration uncertainties. }% of $\pm$20~per~cent.}
\label{fig:NGC2071_mass_compare}
%\end{minipage}
\end{figure*}

\begin{figure*}
%\begin{minipage}{6in}
\centering
\includegraphics[width=5in]{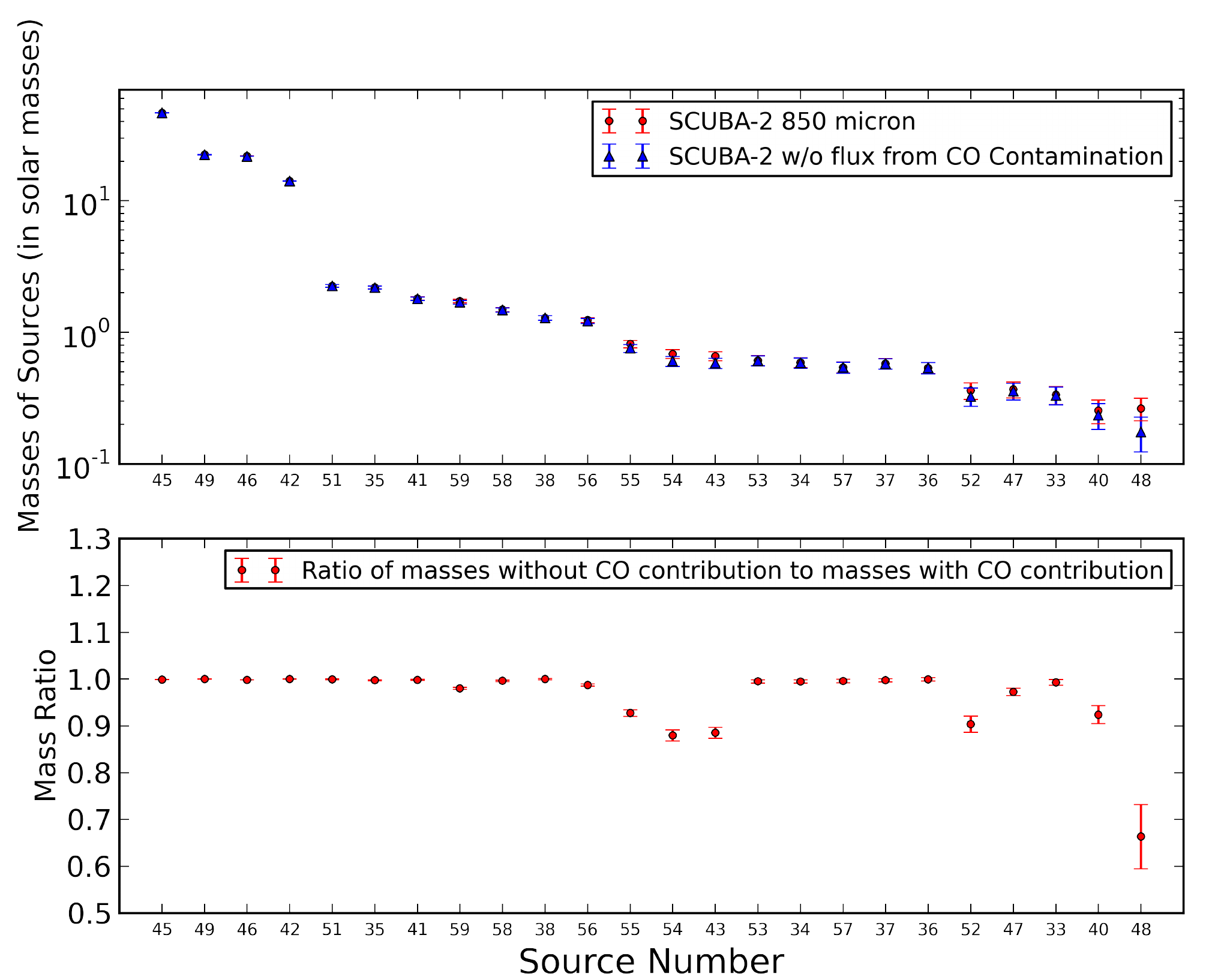}
\caption{{\em{Top:}}  Mass calculations (in solar masses) of the different sources in NGC 2024.  The masses were calculated using the SCUBA-2 850 $\micron$ map and then recalculated excluding the flux contribution from $^{12}$CO in different atmospheric conditions.  {\em{Bottom:}}  Ratio of the masses calculated from the flux without to with the $^{12}$CO contribution.  In both plots, uncertainties are calculated only from the source fluxes and do not include absolute calibration uncertainties.}% of $\pm$20~per~cent.}
\label{fig:NGC2024_mass_compare}
%\end{minipage}
\end{figure*}

Masses were calculated from the source fluxes of the 850~$\micron$ continuum emission (see Section \ref{flux}) with and without $^{12}$CO 3--2 contribution taken into account.   The bottom portion of Figures~\ref{fig:NGC2071_mass_compare} and \ref{fig:NGC2024_mass_compare} show the ratio between these masses for NGC~2071 and NGC~2024 respectively.  Due to the CO flux contamination, the calculated source masses are being overestimated by a factor up to 4.8 for NGC~2071 and 1.5 for NGC~2024.

\subsection{Molecular Outflow Analysis}
\label{outflows}

%{\color{red}{(Aims and motivation paragraph.)}}

The location of protostellar sources can help identify the potential causes of $^{12}$CO contamination.  The presence of protostellar molecular outflows and hot ambient gas from nearby stars results in bright $^{12}$CO emission, making regions with these characteristics rife with CO contamination.  

Sources with high $^{12}$CO contamination were examined in further detail using the HARP data cubes.  In each of the three regions, sources with greater than 20~per~cent contamination were defined as `sources with high contamination.'  For NGC~1333, there were no sources with greater than 20~per~cent contamination, excluding it from this portion of the high contamination study.  In NGC~2071, four sources fulfilled the high contamination criterion and one source in NGC~2024 fulfilled the criterion.  In order to identify the cause of high contamination, the $^{12}$CO spectra were extracted and analysed for molecular outflows. 

For NGC~2071 and NGC~2024, the linewing criterion used to identify a molecular outflow candidate was a linewing above 1.5 K ($T_A ^\ast$) at $\pm$ 4 km s$^{-1}$ from the core velocity, $v_{\mathrm{LSR}}$.  This linewing criterion method follows the method in \citet{2007A&A...472..187H}.  %Core velocities were obtained from C$^{18}$O 3--2 data \citep{2010MNRAS.401..204B}.  
A core velocity of 10~km~s$^{-1}$ was used for all of the sources in NGC~2024 and NGC~2071 based on C$^{18}$O~3--2 data  \citep{2010MNRAS.401..204B}.  Linewing criteria were based on $T_A ^\ast$~RMS values for the regions ($5\sigma$). %detection (NGC~1333 had RMS values $<$~0.1~K and NGC~2071 \&NGC~2024 had RMS values $<$~0.2~K).

The linewing criterion identifies not only protostars driving molecular outflows, but also sources which are contaminated by outflows along the line of sight.  Outflow candidates were identified using the above criteria and examined further to determine if the source or another protostar was the outflow driving source.  Sources with high contamination that were not outflow candidates were further analysed to determine if there were other causes behind the $^{12}$CO contamination, such as a nearby star heating the gas. 

Highly contaminated sources are listed in Table~\ref{moloutflow}.  Table~\ref{moloutflow} includes the region, source number, RA and Dec, flux calculated from aperture photometry in mJy, $^{12}$CO flux contamination in Grade~2 weather in mJy, percentage contamination from $^{12}$CO, core velocity $v_{\mathrm{LSR}}$ in km s$^{-1}$, and the final molecular outflow candidate result.  The outflow naming convention follows \citet{2009A&A...502..139H}, where a `y' is given when an outflow is present and `n' is given when an outflow is not present.  Sources are marked `?' when there is confusion as to the source of the outflow.  In this case, the potential source causing the outflow detection is listed in a footnote.  

\begin{table*}
\begin{minipage}{7in}
\centering
\caption{List of sources categorised with high $^{12}$CO contamination to the 850~$\micron$ dust continuum.}
\begin{tabular}{l c c c c c c c c }
Region & Source & RA & Dec & Flux\footnote[1]{SCUBA-2 fluxes appear to be lower than seen by SCUBA due to the subtraction of large-scale flux by SCUBA-2.} & $^{12}$CO (Grade 2) & Percentage Cont. & $v_{\mathrm{LSR}}$ & Outflow? \\ 
&  & (J2000) & (J2000) & mJy & mJy & (Grade 2) & km s$^{-1}$ & \\ \hline
NGC2071 & 70 &  05:46:57.6 & 00:20:09 & 495 $\pm$ 40 & 224 $\pm$ 6 & 46 $\pm$ 5 & 10.0 & y \\
& 74 &  05:47:01.0 & 00:20:42 & 268 $\pm$ 40 & 196 $\pm$ 6 & 73 $\pm$ 13 & 10.0 & y \\
& 88 &  05:47:06.7 & 00:23:14 & 112 $\pm$ 39 & 38 $\pm$ 3 & 34 $\pm$ 14 & 10.0 & y? \footnote[2]{Could be due to a large, central blue outflow from source at (J2000) 05:47:06.9, 00:22:39 (source 84; LBS-MM19) or a source at (J2000) 05:47:04.1, 00:21:58 (LBS-MM18; NGC2071-IRS), where LBS-MM18 was found to be responsible for driving the outflow in \citet{2001A&A...372L..41M} and both sources are confirmed Class~1 protostars detected using IRAC \citep{2007MNRAS.374.1413N}.\label{fn:repeat}} \\
& 91 &  05:47:08.9 & 00:23:56 & 166 $\pm$ 40 & 132 $\pm$ 5 & 79 $\pm$ 22 & 10.0 & y?$^{\ref{fn:repeat}}$ \\
NGC2024 & 48 & 05:41:19.9 & -01:54:16 & 168 $\pm$ 33 & 56 $\pm$ 3 & 34 $\pm$ 9 & 10.0 & y \\
\hline
\end{tabular}
%\footnotetext[1]{SCUBA-2 fluxes appear to have lower flux than seen by SCUBA due to the subtraction of large-scale flux by SCUBA-2.}
\label{moloutflow}
\end{minipage}
\end{table*}

\subsubsection{Sources in NGC~1333}
\label{ngc_1333_flows}
\begin{figure}
\centering
\includegraphics[width=3.5in]{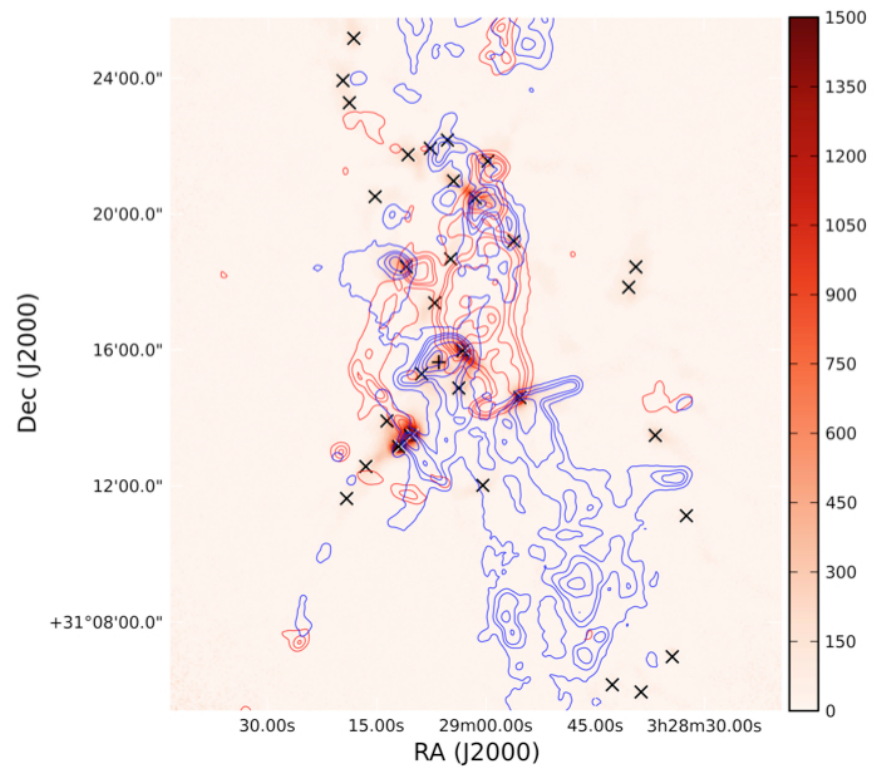}
\caption{SCUBA-2 850~$\micron$ map of NGC~1333.  The colour bar represents flux in mJy~beam$^{-1}$.  Blue contours correspond to blueshifted $^{12}$CO~3--2 HARP intensity $\int{T_A ^\ast} \ d{\mathrm{v}}$ (integrated from -2.5 to 4.5~km~s$^{-1}$).  Red contours correspond to redshifted intensity (integrated from 10.5 to 16.5~km~s$^{-1}$).  Contour levels are 5, 10, 15, 25, 45, 65, and 85~K~km~s$^{-1}$.  Sources in NGC~1333 are denoted by percentage contamination, where `$\times$' denotes sources with 0 to 10~per~cent contamination and `$+$' denotes sources with 10 to 20~per~cent contamination. }% $^{12}$CO~3--2 integrated intensities $\int{T_A ^\ast} \ d{\mathrm{v}}$ in NGC~1333.  The background is an integrated intensity map of $^{12}$CO ranging over 4.5-10.5~km~s$^{-1}$ represents the ambient gas in the region.  Blue contours correspond to blueshifted gas in the cloud using an integrated intensity map ranging over -2.5-4.5~km~s$^{-1}$ with contour levels 5, 10, 15, 25, 45, 65, and 85~K~km~s$^{-1}$.  Red contours correspond to redshifted gas in the cloud using an integrated intensity map ranging over 10.5-16.5~km~s$^{-1}$ with the same contour levels.  Sources with high contamination are labeled with a 15$''$ radius circle.  The colour bar gives the ambient cloud intensity in K~km~s$^{-1}$.}
\label{fig:COcont_molecular}
\end{figure}

Figure \ref{fig:COcont_molecular} shows the 850~$\micron$ SCUBA-2 map with blue contours tracing the blueshifted $^{12}$CO HARP intensity $\int{T_A ^\ast} \ d{\mathrm{v}}$ (integrated from -2.5 to 4.5~km~s$^{-1}$) and red contours tracing the redshifted $^{12}$CO intensity (integrated from 10.5 to 16.5~km~s$^{-1}$).  Sources in NGC~1333 are denoted by the percentage contamination, where `$\times$' denotes sources with 0 to 10~per~cent contamination and `$+$' denotes sources with 10 to 20~per~cent contamination.  %Even though there were no sources exceeding the 20~per~cent criteria, 
Source~21 had the highest percentage contamination at 12~per~cent.  According to the linewing criteria used to identify a molecular outflow candidate (linewing above 1.5 K for $T_A ^\ast$ at $\pm$ 3~km~s$^{-1}$ from the core velocity 7.9~km~s$^{-1}$, following the criterion for NGC~1333 used in \citealt{2007A&A...472..187H}), source~21 is a molecular outflow candidate that could potentially be the result of a source at (J2000) 03:29:03.2, 31:15:59.0 (SVS13) or source at (J2000) 03:29:08.8, 31:15:18.1 (SK-16) \citep{2009A&A...502..139H}.  A $^{12}$CO spectrum for this source is included in Figure~\ref{fig:spectra_sources}.

%Out of the seven sources analysed, the four highest percentages of $^{12}$CO contamination (sources 21, 53, 55, and 62) are due to outflows, determined by the linewing criteria above, where the percent contamination ranged from 20~per~cent-35~per~cent.  When observing the sources based on the measured flux, outflows appear to affect sources of both high and low flux, while non-outflow causes affect sources only $< 530$ mJy/beam.  

\subsubsection{Sources in NGC~2071}
\begin{figure}
\centering
\includegraphics[width=3.5in]{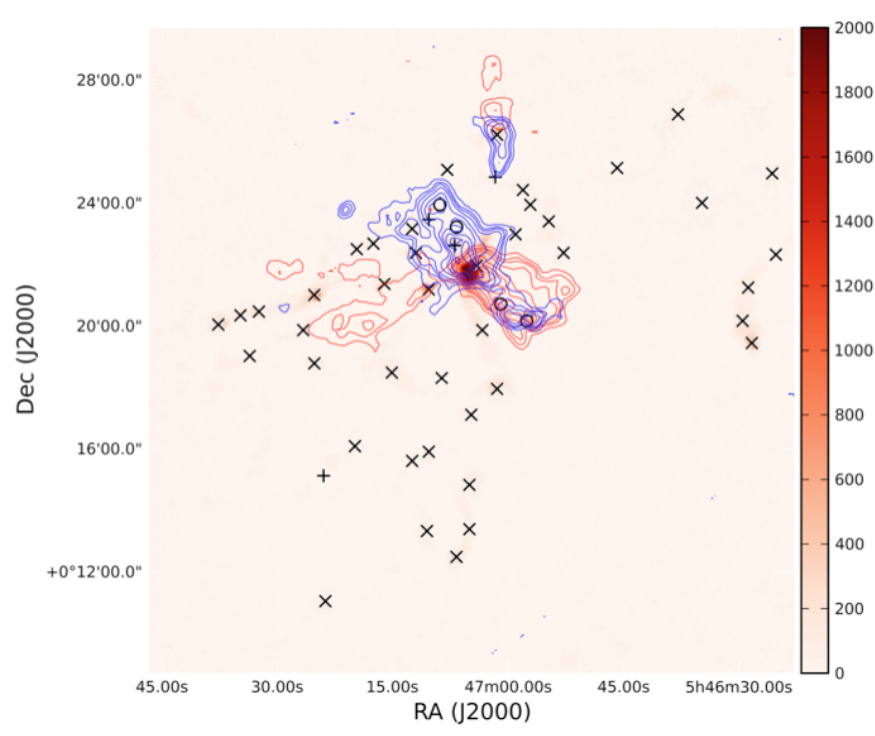}
\caption{SCUBA-2 850~$\micron$ map of NGC~2071.  The colour bar represents flux in mJy~beam$^{-1}$.  Blue contours correspond to blueshifted $^{12}$CO~3--2 HARP intensity $\int{T_A ^\ast} \ d{\mathrm{v}}$ (integrated from -2.0 to 6.0~km~s$^{-1}$).  Red contours correspond to redshifted intensity (integrated from 14.0 to 22.0~km~s$^{-1}$).  Contour levels are 5, 10, 15, 25, 45, 65, 85, 105, 125, and 145~K~km~s$^{-1}$.  Sources in NGC~2071 are denoted by percentage contamination, where `$\times$' denotes sources with 0 to 10~per~cent contamination, `$+$' denotes sources with 10 to 20~per~cent contamination, and `$\mathrm{O}$' denotes sources with greater than 20~per~cent contamination. }%Plot of $^{12}$CO~3--2 integrated intensity $\int{T_A ^\ast} \ d{\mathrm{v}}$ in NGC~2071.  The background is an integrated intensity map of $^{12}$CO ranging from 6.0-14.0~km~s$^{-1}$ to represent the ambient gas in the molecular cloud.  Blue contours correspond to blueshifted regions in the cloud using an integrated intensity map ranging from -2.0-6.0~km~s$^{-1}$ with contour levels 5, 10, 15, 25, 45, 65, 85, 105, 124, and 145~K~km~s$^{-1}$.  Red contours correspond to redshifted regions in the cloud using an integrated intensity map ranging from 14.0-22.0~km~s$^{-1}$ with the same contour levels.  The colour bar gives the ambient cloud intensity in K~km~s$^{-1}$.}
\label{fig:COcont_molecular_orionbn}
\end{figure}

Figure~\ref{fig:COcont_molecular_orionbn} shows 850~$\micron$ SCUBA-2 map with blue contours tracing the blueshifted $^{12}$CO HARP intensity $\int{T_A ^\ast} \ d{\mathrm{v}}$ (integrated from -2.0 to 6.0~km~s$^{-1}$) and red contours tracing the redshifted $^{12}$CO intensity (integrated from 14.0 to 22.0~km~s$^{-1}$).  Sources in NGC~2071 are denoted by the percentage contamination, where `$\times$' denotes sources with 0 to 10~per~cent contamination, `$+$' denotes sources with 10 to 20~per~cent contamination, and `$\mathrm{O}$' denotes sources with greater than 20~per~cent contamination.  According to the linewing criteria used, all four sources with high contamination are molecular outflow candidates.  Even though sources~70 and 74 have clear blue- and redshifted spectral linewings, all four sources appear to trace a large central outflow that could be driven by a source at (J2000) 05:47:06.9, 00:22:39 (source 84; LBS-MM19) or a source at (J2000) 05:47:04.1, 00:21:58 (source 79; LBS-MM18; NGC2071-IRS), where LBS-MM18 was found to be responsible for driving the outflow in \citet{2001A&A...372L..41M} and both sources are confirmed Class~1 protostars detected using IRAC \citep{2007MNRAS.374.1413N}.  Note that sources 84 (LBS-MM19) and 90, both with 18~per~cent contamination, also correlate with the central outflow.  The $^{12}$CO spectra for these sources are displayed in Figure~\ref{fig:spectra_sources}.  

\subsubsection{Sources in NGC~2024}

\begin{figure}
\centering
\includegraphics[width=3.5in]{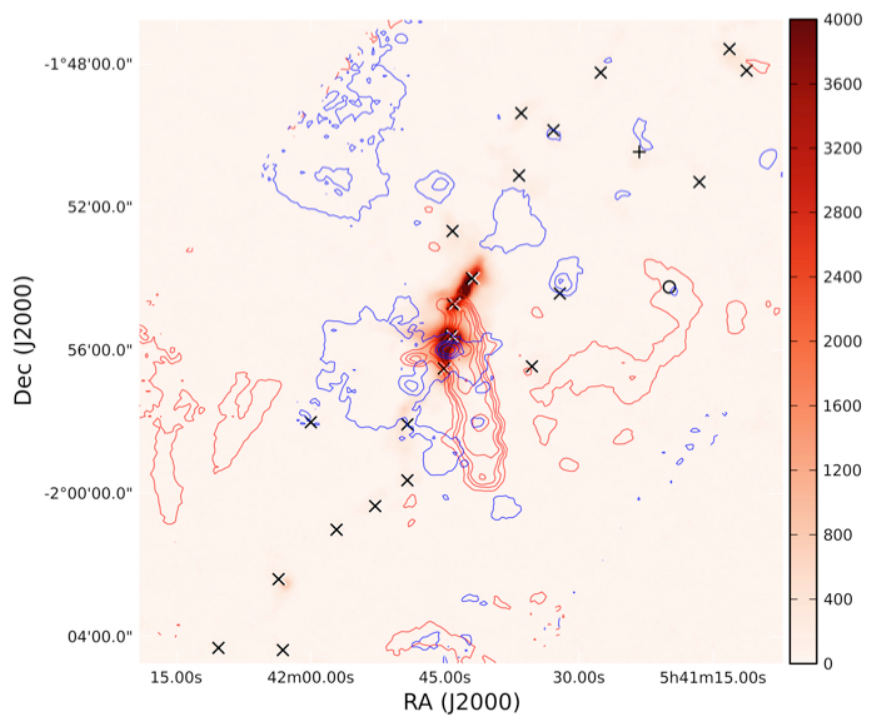}
\caption{SCUBA-2 850~$\micron$ map of NGC~2024.  The colour bar represents flux in mJy~beam$^{-1}$.  Blue contours correspond to blueshifted $^{12}$CO~3--2 HARP intensity $\int{T_A ^\ast} \ d{\mathrm{v}}$ (integrated from -2.0 to 6.0~km~s$^{-1}$).  Red contours correspond to redshifted intensity (integrated from 14.0 to 22.0~km~s$^{-1}$).  Contour levels are 5, 10, 15, 25, 45, 65, 85, 105, 125, and 145~K~km~s$^{-1}$.  Sources in NGC~2024 are denoted by percentage contamination, where `$\times$' denotes sources with 0 to 10~per~cent contamination, `$+$' denotes sources with 10 to 20~per~cent contamination, and `$\mathrm{O}$' denotes sources with greater than 20~per~cent contamination. }%Plot of $^{12}$CO 3--2 integrated intensity $\int{T_A ^\ast} \ d{\mathrm{v}}$ in NGC~2024.  The background is an integrated intensity map of $^{12}$CO ranging from 6.0-14.0~km~s$^{-1}$ to represent the ambient gas in the molecular cloud.  Blue contours correspond to blueshifted regions in the cloud using an integrated intensity map ranging from -2.0-6.0 km~s$^{-1}$ with contour levels 5, 10, 15, 25, 45, 65, 85, 105, 124, and 145~K~km~s$^{-1}$.  Red contours correspond to redshifted regions in the cloud using an integrated intensity map ranging from 14.0-22.0~km~s$^{-1}$ with the same contour levels.  Sources labeled with high contamination are labeled with a 15$''$ radius circle.  The colour bar gives the ambient cloud intensity in K~km~s$^{-1}$.}
\label{fig:COcont_molecular_orionbs}
\end{figure}

 Figure~\ref{fig:COcont_molecular_orionbs} shows 850~$\micron$ SCUBA-2 map with blue contours tracing the blueshifted $^{12}$CO HARP intensity $\int{T_A ^\ast} \ d{\mathrm{v}}$ (integrated from -2.0 to 6.0~km~s$^{-1}$) and red contours tracing the redshifted $^{12}$CO intensity (integrated from 14.0 to 22.0~km~s$^{-1}$).  Sources in NGC~2024 are denoted by the percentage contamination, where `$\times$' denotes sources with 0 to 10~per~cent contamination, `$+$' denotes sources with 10 to 20~per~cent contamination, and `$\mathrm{O}$' denotes sources with greater than 20~per~cent contamination.  According to the linewing criteria used, the single source with a high contamination is a molecular outflow candidate.  The $^{12}$CO spectrum for this source is listed in Figure~\ref{fig:spectra_sources}.

\section{Discussion}
\label{analysis}

%{\color{red}{Absolute levels of CO contamination?}}

%\subsection{Analysis of Combined Sample}
Typical $^{12}$CO contamination levels in the observed SCUBA-2 850~$\micron$ emission from NGC~1333, NGC~2071, and NGC~2024 are under 20~per~cent (this includes 95~per~cent of sources, and 88~per~cent of all sources have under 10~per~cent contamination).  Similar results were found for SCUBA, where \citet{2003A&A...412..157J} suggested that $^{12}$CO line contamination is typically under 10~per~cent for submillimetre sources in Orion and \citet{2000MNRAS.318..952D} suggested contamination was $\sim$10~per~cent near the source V380~Orion~NE.  %Extragalactic sources apparently have $^{12}$CO 3--2 contamination similar to these past studies.  In \citet{2010MNRAS.406.1364L}, the CO contamination level was averaged to be $\sim$20~per~cent and $\sim$25~per~cent in \citet{2004MNRAS.349.1428S}. {\color{red}{Change a few things - corrections.}}

%, where extragalactic contamination from $^{12}$CO 3--2 to the 850 $\mu$m observations has been averaged to be roughly 20~per~cent in Leech et al. (2010) and roughly 25~per~cent in Seaquist et al. (2004) and a study of submillimetre sources in Orion (Johnstone et al. 2003 {

%{\color{red}{Significance of CO contamination correction to submm fluxes and masses compared to other uncertainties (absolute,relative).  When might this be important?  E.G. low mass end of the CMF?  Min column density for SF (Av roughly 8)?  Disk fluxes?  - Ask Jenny P.  Is it possible to estimate the CO contamination based on other factors?  }}

In locations where molecular outflows are present, $^{12}$CO contamination can rise above 20~per~cent and dominate the dust continuum (up to 79~per~cent contamination), corresponding to a CO contribution ranging from 16 to 68~mJy~beam$^{-1}$ for the sources analysed in this study.  Peak $^{12}$CO fluxes found in the molecular outflows of NGC~1333, NGC~2071, and NGC~2024 maps reach even higher fluxes of 84~mJy~beam$^{-1}$, 154~mJy~beam$^{-1}$, and 94~mJy~beam$^{-1}$ respectively.  Our study suggests that molecular outflows can influence line contamination in sources with both high and low continuum fluxes ($\sim$~100 to 500~mJy).  %, while warm molecular gas caused by other factors besides outflows (e.g. nearby stars or large-scale $^{12}$CO flux) only influences sources with fluxes less than 550~mJy.  
This result agrees with that of \citet{2003A&A...412..157J}, where they concluded that the areas with warmer molecular gas temperatures and higher velocities (i.e.\ shocks and molecular outflows) were the only locations where $^{12}$CO emission dominated the dust continuum flux due to the higher molecular line integrated intensities associated with such regions.  In addition, \citet{2003A&A...401L...5G} found a 20~per~cent contamination for the well-known outflow source L1157.  \citet{2009A&A...502..139H} similarly found a 20 to 30~per~cent $^{12}$CO contamination level in the IRAS 03282+3035 outflow in Perseus.  

The $^{12}$CO contamination combined with contamination from other molecular lines allows outflows to potentially be seen in continuum maps with a similar appearance as protostellar cores or filamentary structure, which may be the case for SVS13 in NGC~1333 and the large, central outflows in NGC~2071 and NGC~2024 that have strong evidence of molecular outflow lobes detected in the dust continuum emission maps (see Figure~\ref{fig:codust_ngc1333}).  Other studies have suggested that regions involving molecular outflows can reach 50~per~cent $^{12}$CO contamination, e.g.\ in the extended outflow lobes of V380 Orion NE \citep{2000MNRAS.318..952D}, and even up to 100~per~cent contamination, e.g. \ the central blue outflow region in NGC~2071 driven by source LBS-MM18 (NGC2071-IRS;  \citealt{2001A&A...372L..41M}).  %molecular line contribution (using $^{12}$CO~3--2 and HCO$^{+}$~4--3 maps).  

%In locations where molecular outflows are not present and the peak temperature of the $^{12}$CO spectra is low (typically $T_\mathrm{MB} < 20$~K), the $^{12}$CO can still contribute up to a flux density of 75~mJy, or 25~mJy~beam$^{-1}$.  It is not clear if small-scale CO structure contributes to the SCUBA-2 flux in this case or if the S2SRO insensitivity to large-scale structure is not fully matched by the GSM filter.  Molecular line contamination at these low levels can be a particular problem for sources with low flux.  For example, in NGC~2071, the four sources with the lowest flux ($< $~80~mJy in sources 56, 62, 97, and 98) have a $^{12}$CO contamination of 26~per~cent-111~per~cent (where source~56 has $>$~100~per~cent contamination).  In addition to source~56, the other five NGC~2071 sources (source~70, 74, 88, 90 and 91) and one NGC~2024 source (source 48) that have higher flux contribution from the $^{12}$CO than the 850~$\mu$m dust emission point to the need for a more detailed model of the SCUBA-2 structure response.  This is something we will consider for the full SCUBA-2 array, but not for the limited S2SRO data.

The FCF uncertainty from calibrator observations for the 850~$\micron$ S2SRO maps is 18~per~cent (SMURF SCUBA-2 Data Reduction Cookbook)\footnote{\url{http://star-www.rl.ac.uk/star/docs/sc19.htx/node40.html}}.  The calibration uncertainty of HARP observations at JCMT is estimated to be 20~per~cent by \citet{2009MNRAS.399.1026B}.  With contamination levels to SCUBA-2~850~$\micron$ less than or equal to 20~per~cent for the majority of the sources, the typical contamination is less than or equivalent to the calibration uncertainty.  Problems arise when the contamination is greater than than calibration uncertainties, contributing a significant portion of flux and potentially dominating the dust continuum.  For bright sources, it should be possible to use the subtracted $^{12}$CO background to estimate the column density and hence the potential CO contamination to the 850~$\micron$ SCUBA-2 dust emission, %increase in 850~$\micron$ dust emission to SCUBA-2 due to CO contamination, %correct the contamination problem by mapping the regions in $^{12}$CO in order to subtract the line flux contribution from the dust continuum directly, 
as suggested by \citet{2002ApJ...580..285T}.  

For faint sources, the insensitivity of SCUBA-2 to the large-scale dust emission introduces additional uncertainties.  The spatially filtered maps created here do not appear to entirely subtract the total large-scale flux detected by HARP.  Excess $^{12}$CO flux seen as positive flux in residuals (Section~\ref{GSM}) point to the need for a more detailed model of SCUBA-2 structure response.  The large-scale reconstruction issues limit analysing contamination where the SCUBA-2 flux is faint.  This analysis is something we will consider for the full SCUBA-2 array, but not for the limited S2SRO data.  
%For faint sources, the insensitivity of SCUBA-2 to the large-scale dust emission introduces additional uncertainties.  %However, this process is complicated by large-scale flux seen in the HARP $^{12}$CO~3--2 maps that is not detected by SCUBA-2. 
%The spatially filtered maps created here do not appear to entirely subtract the total large-scale flux detected by HARP, indicated by the positive flux caused by high $^{12}$CO flux in the $1'$ GSM residuals of the regions.  %NGC~2071 and NGC~2024 when several sources had a higher molecular line flux than a dust emission flux.  
If the SCUBA-2 and HARP maps were subtracted in order to account for the molecular line contamination, there is a possibility that an overcompensation for the molecular line flux would occur, creating regions of negative flux in the dust continuum map.  

An additional uncertainty in the contamination due to large-scale CO emission is the inclusion of the secondary beam in the calculation of the telescope beam area in Equation~\ref{eq:conversion}, which increases the beam area by a factor of 1.2 at 850~$\micron$ (determined from a new FWHM calculated in Section~\ref{beamarea}).   Large-scale emission couples to the telescope beam as the efficiency factor $\eta_{\mathrm{fss}}$ instead of $\eta_{\mathrm{MB}}$ (as in Equation~\ref{eq:beamcontribution}).  Therefore, the conversion factor for $^{12}$CO~3--2 would increase by a factor of 1.1, which is insignificant compared to other uncertainties.  The scales of large-scale emission that could cause significant signal without being taken out by the S2SRO common-mode subtraction range from $13.8''$ (the FWHM of the 850~$\micron$ beam) to $1'$ (the Gaussian FWHM used in the GSM masking process).  For the full SCUBA-2 array, CO emission on scales up to 8$'$ in size could contribute.  

CO is not the only possible contributor in the 850 and 450~$\micron$ bands.  Studies of other molecular lines found SCUBA 850~$\micron$ line contamination from HCN, HNC, CN and methanol add together to form roughly 40~per~cent of the total line contamination when observing other more energetic sources, like the shocked region SK1-OMC3 \citep{2003A&A...412..157J} .  Similar contamination was found in the Kleinmann-Low nebula, %an energetic region of Orion 
from SO and SO$_2$ emission that was 28 to 50~per~cent of the total line contamination at 850~$\micron$ \citep{1995ApJ...451..238S, 1994ApJS...94..147G}.  Other studies have found the total line contamination by other molecular lines to be a factor of 2 to 3 times that from CO in outflows \citep{2002ApJ...580..285T, 2003A&A...401L...5G}.  Since molecular line contamination from other molecules is also likely, some features with low flux in the dust continuum may entirely be the result of line emission.

\subsection{$^{12}$CO~6--5 Contamination}
For the 450~$\micron$ band, we have no $^{12}$CO~6--5 maps with which to estimate the CO contamination directly.  Using the line intensities from the $^{12}$CO~3--2 HARP maps, we can instead %estimate the $^{12}$CO~6--5 intensities in order to 
predict the potential line contamination from $^{12}$CO~6--5 to the SCUBA-2 450~$\micron$ dust continuum signal.  %{\color{red}{Importance?  Location?  Regions with higher temperature?}}

%Molecular line intensity is directly related to the brightness temperature, T$_{\mathrm{b}}$.  
Assuming local thermodynamic equilibrium, we can estimate the ratio of the main-beam brightness temperatures T$_{\mathrm{MB}}$ for $^{12}$CO~6--5 and $^{12}$CO~3--2.  We assume the excitation temperature, $T_{\mathrm{ex}}$, is equal to the kinetic temperature of the region, and is therefore the same for both $^{12}$CO~3--2 and $^{12}$CO~6--5.  We also assume the partition function $Z \approx {2 T_{\mathrm{ex}}} / {T_0}$ and the Gaussian line shape $\theta(\mathrm{peak})$~ $=$~${2 c \ \sqrt{2 \ln{2}}} \ / \ {\nu \ \Delta \mathrm{v} \sqrt{2\pi}}$, yielding the relation (in CGS units):

\begin{eqnarray}
T_{\mathrm{MB}} &=& \frac{8 \pi^3}{3 h} \mu^2 (J+1)^2 \frac{T_0 ^2}{2 T_{\mathrm{ex}}} \exp \left({\frac{-(J+1)(J+2) T_0}{2 T_{\mathrm{ex}}}}\right) \nonumber \\ 
&\times& \frac{2 \sqrt{2 \ln 2}}{\Delta \mathrm{v} \sqrt{2 \pi}}
\label{eq:brightnesstemp}
\end{eqnarray} where $h$ is Planck's constant, $\mu$ is the permanent electric dipole moment of the molecule, $J$ is the lower rotational level of a linear molecule, and $T_0$ is the ground-state temperature ($h \nu_0 / k $) at 5.5~K.  Using Equation \ref{eq:brightnesstemp}, the ratio $^{12}$CO~6--5/$^{12}$CO~3--2 is:

\begin{eqnarray}
\frac{T_{\mathrm{MB}} (6 \rightarrow 5)}{T_{\mathrm{MB}} (3 \rightarrow 2)} &=& \frac{(6)^2}{(3)^2} \frac{\exp \left(\frac{- 21 \ T_0}{T_\mathrm{ex}}\right)}{\exp \left(\frac{- 6 \ T_0}{T_{\mathrm{ex}}}\right)} \nonumber \\
&=& 4 \exp {\left(-15 \frac{T_0}{T_\mathrm{ex}} \right) }
\label{eq:cocompare}
\end{eqnarray} where $J(6\rightarrow5) = 5$ for $T_{\mathrm{MB}} (6\rightarrow5)$ and $J(3\rightarrow2) = 2$ for $T_{\mathrm{MB}} (3\rightarrow2)$.

%\begin{equation}
%\frac{T_{\mathrm{MB}} (6 \rightarrow 5)}{T_{\mathrm{MB}} (3 \rightarrow 2)} = \frac{(J(6\rightarrow5) + 1)^2 e^{\frac{- (J(6\rightarrow5)+1) (J(6\rightarrow5)+2) T_o}{2 T_{\mathrm{ex}}}}}{(J(3\rightarrow2) + 1)^2 e^{\frac{- (J(3\rightarrow2)+1) (J(3\rightarrow2)+2) T_o}{2 T_{\mathrm{ex}}}}} 
%\end{equation} where $J(6\rightarrow5) = 5$ for $T_{\mathrm{MB}} (6\rightarrow5)$ and $J(3\rightarrow2) = 2$ for $T_{\mathrm{MB}} (3\rightarrow2)$.

 Assuming the source dust temperatures of 10~K, as in Section \ref{mass}, are equal to the excitation temperature in the protostellar envelope, it follows from Equation~\ref{eq:cocompare} the ratio $^{12}$CO~6--5/$^{12}$CO~3--2 is $\sim$0.001 in the optically thin case.  The ratio between the two lines is low due to the low temperature of the region, indicating there is less likelihood of detecting $^{12}$CO~6--5 in cooler regions of the cloud.  The sources analysed in NGC~1333, NGC~2071, and NGC~2024 with high $^{12}$CO~3--2 contamination correspond to molecular clouds at temperatures of 20 to 25~K.  At 25~K, the ratio of $^{12}$CO~6--5/$^{12}$CO~3--2 is 0.147 in the optically thin case.  However, outflows can contain even higher temperatures, ranging from 50 to 150~K \citep{1999A&A...344..687H, 2009A&A...507.1425V}.  At 50~K, the ratio is 0.769, indicating $^{12}$CO~6--5 is much more likely to be detected from outflows.  If, on the other hand, both lines are optically thick, then the ratio tends to 1 as is known to be the case for $^{12}$CO~3--2 \citep{2010MNRAS.401..204B, 2010MNRAS.408.1516C}.
 
% If, on the other hand, both lines are optically thick (as is known to be the case for $^{12}$CO~3--2 from \citet{2010MNRAS.401..204B, 2010MNRAS.408.1516C}), then the ratio tends to 1.  %that it is more likely to observe $^{12}$CO~6--5 in regions of outflows than in the cooler regions of the ambient cloud material.  

Using the ratio $^{12}$CO~6--5/$^{12}$CO~3--2, we can estimate typical peak fluxes for $^{12}$CO~6--5:  an excitation temperature of 25~K and a source with a typical $^{12}$CO~3--2 integrated intensity of 100~K~km~s$^{-1}$ will produce a corresponding $^{12}$CO~6--5 flux contribution of 8~mJy~beam$^{-1}$ for Grade~2 weather in the 450~$\micron$ SCUBA-2 map.  With the SCUBA-2 450~$\micron$ sensitivity for the Gould Belt Survey \citep{2007PASP..119..855W} at a RMS of $\sim$35~mJy~beam$^{-1}$ for Grade 2 weather,  the $^{12}$CO~6--5 flux contribution would not be detected.  Even in the case of optically thick emission where the ratio $^{12}$CO~6--5/$^{12}$CO~3--2 is 1, the $^{12}$CO~6--5 is estimated to be 57~mJy~beam$^{-1}$, which is under the 5$\sigma$ detection limit.  %{\color{red}{Finish with estimated temperatures of protostellar cores.}}

Using published observations, $^{12}$CO~6--5 contamination can be studied in further detail.  $^{12}$CO~6--5 data were taken for IRAS 2A, 4A, and 4B in NGC~1333 by \citet{2010A&A...521L..40Y}.  These sources are particularly bright and possibly intermediate-mass protostars.  $^{12}$CO~6--5 integrated intensities at the positions of the protostars  were 57~K~km~s$^{-1}$, 122~K~km~s$^{-1}$, and 43~K~km~s$^{-1}$ respectively.  Using the $^{12}$CO~6--5 conversion factors calculated in this study for Grade~2 weather, the corresponding CO contamination to the SCUBA-2~450~$\micron$ dust continuum would be 32~mJy~beam$^{-1}$ (IRAS~2A), 70~mJy~beam$^{-1}$ (IRAS~4A), and 25~mJy~beam$^{-1}$ (IRAS~4B) for the CO contamination to the SCUBA-2 450~$\micron$ dust continuum.  Peak fluxes from SCUBA were 2355~mJy~beam$^{-1}$,  7000~mJy~beam$^{-1}$, and 3025~mJy~beam$^{-1}$ respectively \citep{2005A&A...440..151H}.  The SCUBA-2~450~$\micron$ dust emission peaks are a factor of several hundred times larger than the $^{12}$CO~6--5 contribution.  For these bright protostars, the CO contamination is insignificant at 450~$\micron$.

Dust continuum fluxes in the 450~$\micron$ SCUBA-2 band are estimated to be a factor of 6 to 12 higher than fluxes in the 850~$\micron$ band.  In the Rayleigh-Jeans approximation, the dust optical depth increases as $\lambda^{-\beta}$ with $\beta$ between 1 and 2 and the corresponding flux density increases as $\lambda^{-3}$ to $\lambda^{-4}$ \citep{2009arXiv0903.0562W}.  The increase in continuum flux at 450~$\micron$ clearly outweighs the expected contribution from $^{12}$CO~6--5.  At most the $^{12}$CO~6--5 integrated main-beam emission is the same as the $^{12}$CO~3--2 emission assuming optically thick emission.  This situation corresponds to a molecular line ratio of 1 and a contamination flux ratio of 0.84 (assuming Grade~2 weather).  Since 450~$\micron$ continuum fluxes increase with respect to the 850~$\micron$ continuum, 450~$\micron$ continuum measurements would more likely be contaminated by strong molecular outflows instead of other means, i.e.\ nearby stars or ambient cloud emission.  Nonetheless, potential contamination could occur in particularly low flux sources with nearby molecular outflows, such as the sources in NGC~2071 that were near to confirmed outflow candidates (see Section \ref{outflows}).  %({\color{red}{Include depth of maps at at 5 sigma detection, what CO should be to affect SCUBA-2 maps.}})

In the case of large-scale $^{12}$CO~6--5 emission, it is also necessary to include the secondary beam in the calculation of the 450~$\micron$ telescope beam (Equation~\ref{eq:conversion}).  The total beam area for 450~$\micron$ increases by a factor of 2.0 (determined from an effective FWHM calculated in Section~\ref{beamarea}).  Using Equation~\ref{eq:beamcontribution} with the the efficiency factor $\eta_{\mathrm{fss}}$, the $^{12}$CO~6--5 conversion factor would increase by a factor of 1.7.  The increase in the expected 450~$\micron$ dust continuum flux still exceeds any change in the $^{12}$CO line conversion factors.  %    The increase in the $^{12}$CO~6--5 conversion factors is still less than the increase in the expected dust continuum flux for the 450~$\micron$ maps.  
Therefore, our conclusion that there is little CO contamination in the 450~$\micron$ maps, as discussed above, still holds.

\section{Conclusions}
\label{conclusions}

In this study, the $^{12}$CO line contamination factors for the 450~$\micron$ and 850~$\micron$ SCUBA-2 continuum bands were calculated under different atmospheric conditions (weather grades~1 to 5).  These contamination factors were then applied to three different regions, NGC~1333, NGC~2071, and NGC~2024, in order to study the HARP $^{12}$CO~3--2 flux contribution to the SCUBA-2 850~$\micron$ measurements using a list of sources for each region.  Sources with high $^{12}$CO contamination (greater than 20~per~cent) were analysed in further detail to determine the cause of the contamination.  The following can be concluded from this study:

\begin{enumerate}
	\item For the 850~$\micron$ SCUBA-2 filter profile, the $^{12}$CO~3--2 contamination factors increase as the sky opacity $\tau_{225}$ increases.  The contamination factors (mJy~beam$^{-1}$~per~K~km~s$^{-1}$) of $^{12}$CO to the 850~$\micron$ dust emission are, by weather grade,  (Grade~1)~0.63; (Grade~2)~0.68; (Grade~3)~0.70; (Grade~4)~0.74; (Grade~5)~0.77.  
	\item For the 450~$\micron$ SCUBA-2 filter profile, the $^{12}$CO~6--5 contamination factors decrease as the sky opacity $\tau_{225}$ increases due to the atmosphere transmission steeply declining at higher opacity grades.  The contribution factors (mJy~beam$^{-1}$~per~K~km~s$^{-1}$) of $^{12}$CO to the 450~$\micron$ dust emission are, by weather grade,  (Grade~1)~0.64; (Grade~2)~0.57; (Grade~3)~0.51; (Grade~4)~0.41; (Grade~5)~0.35.
	\item The $^{12}$CO~3--2 contribution to the 850~$\micron$ SCUBA-2 dust continuum is typically under 20~per~cent %(a source with a $\sim$450~mJy dust continuum flux with a 20~per~cent contamination would have corresponding $^{12}$CO~3--2 flux of 90~mJy) in 
for all of the regions studied.  However, in regions of molecular outflows, the $^{12}$CO can reach a flux contribution of $\sim$~68~mJy~beam$^{-1}$ for the sources studied, dominating the dust continuum in sources with both high and low continuum flux densities (up to 500~mJy~beam$^{-1}$) with a contribution up to 79~per~cent contamination.  Peak $^{12}$CO fluxes in molecular outflows in the regions reached even higher levels, up to 154~mJy~beam$^{-1}$.  There is strong evidence that $^{12}$CO~3--2 contamination, while mostly minimal, is a major potential source of confusion that can be observed directly in the 850~$\micron$ dust continuum maps resembling protostellar cores or filamentary structure (as may be the case in NGC~1333, NGC~2071, and NGC~2024).%  In regions of hot molecular gas, e.g.\ heated by nearby O and B stars,$^{12}$CO contribution to the dust continuum can also dominate the flux, but this only occurs in sources with a flux less than  550~mJy.  Lastly, there are regions where outflows and nearby stars do not seem to be the cause of the $^{12}$CO~3--2 contamination.  In these regions, $^{12}$CO can contribute up to $\sim$25~mJy~beam$^{-1}$.  In sources with low flux, this contribution can be particularly strong.  This contribution comes from ambient molecular gas.  At these levels, the exact CO contamination is uncertain due to the uncharacterised coupling of SCUBA-2 to large-scale emission.  This is further supported by the six sources in NGC~2071 (sources~56, 70, 74, 88, 90 and 91) and one source in NGC~2024 (source 48) with a $^{12}$CO flux contribution higher than the SCUBA-2 850~$\micron$ dust continuum fluxes.
	\item Even though we have no $^{12}$CO~6--5 molecular line maps to study in further detail, in hot (50~K) regions, e.g.\ molecular outflows, the ratio of main-beam temperature $T_\mathrm{MB}$ for $^{12}$CO~6--5/$^{12}$CO~3--2 is $\sim$0.769.  %With the ratio of  $^{12}$CO~6--5/$^{12}$CO~3--2 at 0.147 for molecular cloud temperatures of 20 to 25~K and a ratio of $\sim$0.001 for temperatures of 10~K in the protostellar envelope the source of $^{12}$CO~6--5 contamination would be most likely due to molecular outflows.  
However, CO contamination to the 450~$\micron$ source fluxes is not expected to be as much of an issue because of the expectation for the 450~$\micron$ dust emission to be a factor of 6 to 12~times brighter than the 850~$\micron$ fluxes.    %However, the 450~$\micron$ dust emission is estimated to be a factor of 6-12 times brighter than 850~$\micron$ fluxes.  Therefore, CO contamination to the 450~$\micron$ source fluxes is not expected to be as much of an issue.  %{\color{red}{Explain the $^{12}$CO~6--5 line contamination possibilities in the 450~$\micron$ maps.}}
\end{enumerate}

\section{Acknowledgments}
We would like to thank Antonio Chrysostomou and Simon Coud\'e for their helpful input for this paper.  The JCMT is operated by the Joint Astronomy Centre (JAC) on behalf of the Science and Technology Facilities Council (STFC) of the United Kingdom, the National Research Council of Canada, and the Netherlands Organisation for Scientific Research.  This work made use of SIMBAD that is operated at CDS, Strasbourg, France.  We acknowledge the data analysis facilities provided by the Starlink Project which is run by CCLRC on behalf of PPARC. In addition, the following Starlink package AUTOPHOTOM has been used.  This research made use of APLpy, an open-source plotting package for Python hosted at \url{http://aplpy.github.com}.  ED acknowledges the support of a college studentship from the University of Exeter.  

\bibliography{cocont}	

\begin{thebibliography}{}

\bibitem[\protect\citeauthoryear{{Aguirre}, {Ginsburg}, {Dunham}, {Drosback},
  {Bally}, {Battersby}, {Bradley}, {Cyganowski}, {Dowell}, {Evans} II, {Glenn},
  {Harvey}, {Rosolowsky}, {Stringfellow}, {Walawender} \& {Williams}}{{Aguirre}
  et~al.}{2011}]{2011ApJS..192....4A}
{Aguirre} J.~E.,  {Ginsburg} A.~G.,  {Dunham} M.~K.,  {Drosback} M.~M.,
  {Bally} J.,  {Battersby} C.,  {Bradley} E.~T.,  {Cyganowski} C.,  {Dowell}
  D.,  {Evans} II N.~J.,  {Glenn} J.,  {Harvey} P.,  {Rosolowsky} E.,
  {Stringfellow} G.~S.,  {Walawender} J.,    {Williams} J.~P.,  2011, ApJ, 192,
  4

\bibitem[\protect\citeauthoryear{{Andr{\'e}}, {Bouwman}, {Belloche} \&
  {Hennebelle}}{{Andr{\'e}} et~al.}{2003}]{2003cdsf.conf..127A}
{Andr{\'e}} P.,  {Bouwman} J.,  {Belloche} A.,    {Hennebelle} P.,  2003, in
  {C.~L.~Curry \& M.~Fich} ed., SFChem 2002: Chemistry as a Diagnostic of Star
  Formation {Submillimeter Studies of Prestellar Cores and Protostars: Probing
  the Initial Conditions for Protostellar Collapse}.
pp 127--+

\bibitem[\protect\citeauthoryear{{Andre}, {Ward-Thompson} \& {Motte}}{{Andre}
  et~al.}{1996}]{1996A&A...314..625A}
{Andre} P.,  {Ward-Thompson} D.,    {Motte} F.,  1996, A\&A, 314, 625

\bibitem[\protect\citeauthoryear{{Brown}, {de Geus} \& {de Zeeuw}}{{Brown}
  et~al.}{1994}]{1994A&A...289..101B}
{Brown} A.~G.~A.,  {de Geus} E.~J.,    {de Zeeuw} P.~T.,  1994, A\&A, 289, 101

\bibitem[\protect\citeauthoryear{{Buckle}, {Curtis}, {Roberts}, {White} \&
  {Hatchell}}{{Buckle} et~al.}{2010}]{2010MNRAS.401..204B}
{Buckle} J.~V.,  {Curtis} E.~I.,  {Roberts} J.~F.,  {White} G.~J.,
  {Hatchell} J.,  2010, MNRAS, 401, 204

\bibitem[\protect\citeauthoryear{{Buckle}, {Hills}, {Smith}, {Dent} \&
  {Bell}}{{Buckle} et~al.}{2009}]{2009MNRAS.399.1026B}
{Buckle} J.~V.,  {Hills} R.~E.,  {Smith} H.,  {Dent} W.~R.~F.,    {Bell} G.,
  2009, MNRAS, 399, 1026

\bibitem[\protect\citeauthoryear{{Curtis}, {Richer}, {Swift} \&
  {Williams}}{{Curtis} et~al.}{2010}]{2010MNRAS.408.1516C}
{Curtis} E.~I.,  {Richer} J.~S.,  {Swift} J.~J.,    {Williams} J.~P.,  2010,
  MNRAS, 408, 1516

\bibitem[\protect\citeauthoryear{{Davis}, {Dent}, {Matthews}, {Coulson} \&
  {McCaughrean}}{{Davis} et~al.}{2000}]{2000MNRAS.318..952D}
{Davis} C.~J.,  {Dent} W.~R.~F.,  {Matthews} H.~E.,  {Coulson} I.~M.,
  {McCaughrean} M.~J.,  2000, MNRAS, 318, 952

\bibitem[\protect\citeauthoryear{{Di Francesco}, {Johnstone}, {Kirk},
  {MacKenzie} \& {Ledwosinska}}{{Di Francesco}
  et~al.}{2008}]{2008ApJS..175..277D}
{Di Francesco} J.,  {Johnstone} D.,  {Kirk} H.,  {MacKenzie} T.,
  {Ledwosinska} E.,  2008, ApJS, 175, 277

\bibitem[\protect\citeauthoryear{{Enoch}, {Young}, {Glenn}, {Evans} II,
  {Golwala}, {Sargent}, {Harvey}, {Aguirre}, {Goldin}, {Haig}, {Huard},
  {Lange}, {Laurent}, {Maloney}, {Mauskopf}, {Rossinot} \& {Sayers}}{{Enoch}
  et~al.}{2006}]{2006ApJ...638..293E}
{Enoch} M.~L.,  {Young} K.~E.,  {Glenn} J.,  {Evans} II N.~J.,  {Golwala} S.,
  {Sargent} A.~I.,  {Harvey} P.,  {Aguirre} J.,  {Goldin} A.,  {Haig} D.,
  {Huard} T.~L.,  {Lange} A.,  {Laurent} G.,  {Maloney} P.,  {Mauskopf} P.,
  {Rossinot} P.,    {Sayers} J.,  2006, ApJ, 638, 293

\bibitem[\protect\citeauthoryear{{Evans} II, {Rawlings}, {Shirley} \&
  {Mundy}}{{Evans} et~al.}{2001}]{2001ApJ...557..193E}
{Evans} II N.~J.,  {Rawlings} J.~M.~C.,  {Shirley} Y.~L.,    {Mundy} L.~G.,
  2001, ApJ, 557, 193

\bibitem[\protect\citeauthoryear{{Gordon}}{{Gordon}}{1995}]{1995A&A...301..853G}
{Gordon} M.~A.,  1995, A\&A, 301, 853

\bibitem[\protect\citeauthoryear{{Groesbeck}, {Phillips} \&
  {Blake}}{{Groesbeck} et~al.}{1994}]{1994ApJS...94..147G}
{Groesbeck} T.~D.,  {Phillips} T.~G.,    {Blake} G.~A.,  1994, ApJS, 94, 147

\bibitem[\protect\citeauthoryear{{Gueth}, {Bachiller} \& {Tafalla}}{{Gueth}
  et~al.}{2003}]{2003A&A...401L...5G}
{Gueth} F.,  {Bachiller} R.,    {Tafalla} M.,  2003, A\&A, 401, L5

\bibitem[\protect\citeauthoryear{{Hatchell} \& {Dunham}}{{Hatchell} \&
  {Dunham}}{2009}]{2009A&A...502..139H}
{Hatchell} J.,  {Dunham} M.~M.,  2009, A\&A, 502, 139

\bibitem[\protect\citeauthoryear{{Hatchell}, {Fuller} \& {Ladd}}{{Hatchell}
  et~al.}{1999}]{1999A&A...344..687H}
{Hatchell} J.,  {Fuller} G.~A.,    {Ladd} E.~F.,  1999, A\&A, 344, 687

\bibitem[\protect\citeauthoryear{{Hatchell}, {Fuller} \& {Richer}}{{Hatchell}
  et~al.}{2007}]{2007A&A...472..187H}
{Hatchell} J.,  {Fuller} G.~A.,    {Richer} J.~S.,  2007, A\&A, 472, 187

\bibitem[\protect\citeauthoryear{{Hatchell}, {Fuller}, {Richer}, {Harries} \&
  {Ladd}}{{Hatchell} et~al.}{2007}]{2007A&A...468.1009H}
{Hatchell} J.,  {Fuller} G.~A.,  {Richer} J.~S.,  {Harries} T.~J.,    {Ladd}
  E.~F.,  2007, A\&A, 468, 1009

\bibitem[\protect\citeauthoryear{{Hatchell}, {Richer}, {Fuller}, {Qualtrough},
  {Ladd} \& {Chandler}}{{Hatchell} et~al.}{2005}]{2005A&A...440..151H}
{Hatchell} J.,  {Richer} J.~S.,  {Fuller} G.~A.,  {Qualtrough} C.~J.,  {Ladd}
  E.~F.,    {Chandler} C.~J.,  2005, A\&A, 440, 151

\bibitem[\protect\citeauthoryear{{Hildebrand}}{{Hildebrand}}{1983}]{1983QJRAS..24..267H}
{Hildebrand} R.~H.,  1983, QJRAS, 24, 267

\bibitem[\protect\citeauthoryear{{Holland}, {Duncan} \& {Griffin}}{{Holland}
  et~al.}{2002}]{2002ASPC..278..463H}
{Holland} W.,  {Duncan} W.,    {Griffin} M.,  2002, in {S.~Stanimirovic,
  D.~Altschuler, P.~Goldsmith, \& C.~Salter} ed., Single-Dish Radio Astronomy:
  Techniques and Applications Vol.~278 of Astronomical Society of the Pacific
  Conference Series, {Bolometers for Submillimeter and Millimeter Astronomy}.
pp 463--491

\bibitem[\protect\citeauthoryear{{Johnstone} \& {Bally}}{{Johnstone} \&
  {Bally}}{1999}]{1999ApJ...510L..49J}
{Johnstone} D.,  {Bally} J.,  1999, ApJL, 510, L49

\bibitem[\protect\citeauthoryear{{Johnstone} \& {Bally}}{{Johnstone} \&
  {Bally}}{2006}]{2006ApJ...653..383J}
{Johnstone} D.,  {Bally} J.,  2006, ApJ, 653, 383

\bibitem[\protect\citeauthoryear{{Johnstone}, {Boonman} \& {van
  Dishoeck}}{{Johnstone} et~al.}{2003}]{2003A&A...412..157J}
{Johnstone} D.,  {Boonman} A.~M.~S.,    {van Dishoeck} E.~F.,  2003, A\&A, 412,
  157

\bibitem[\protect\citeauthoryear{{Johnstone}, {Wilson}, {Moriarty-Schieven},
  {Joncas}, {Smith}, {Gregersen} \& {Fich}}{{Johnstone}
  et~al.}{2000}]{2000ApJ...545..327J}
{Johnstone} D.,  {Wilson} C.~D.,  {Moriarty-Schieven} G.,  {Joncas} G.,
  {Smith} G.,  {Gregersen} E.,    {Fich} M.,  2000, ApJ, 545, 327

\bibitem[\protect\citeauthoryear{{Kirk}, {Johnstone} \& {Di Francesco}}{{Kirk}
  et~al.}{2006}]{2006ApJ...646.1009K}
{Kirk} H.,  {Johnstone} D.,    {Di Francesco} J.,  2006, ApJ, 646, 1009

\bibitem[\protect\citeauthoryear{{Lada}, {Alves} \& {Lada}}{{Lada}
  et~al.}{1996}]{1996AJ....111.1964L}
{Lada} C.~J.,  {Alves} J.,    {Lada} E.~A.,  1996, AJ, 111, 1964

\bibitem[\protect\citeauthoryear{{Launhardt}, {Mezger}, {Haslam}, {Kreysa},
  {Lemke}, {Sievers} \& {Zylka}}{{Launhardt}
  et~al.}{1996}]{1996A&A...312..569L}
{Launhardt} R.,  {Mezger} P.~G.,  {Haslam} C.~G.~T.,  {Kreysa} E.,  {Lemke} R.,
   {Sievers} A.,    {Zylka} R.,  1996, A\&A, 312, 569

\bibitem[\protect\citeauthoryear{{Mitchell}, {Johnstone}, {Moriarty-Schieven},
  {Fich} \& {Tothill}}{{Mitchell} et~al.}{2001}]{2001ApJ...556..215M}
{Mitchell} G.~F.,  {Johnstone} D.,  {Moriarty-Schieven} G.,  {Fich} M.,
  {Tothill} N.~F.~H.,  2001, ApJ, 556, 215

\bibitem[\protect\citeauthoryear{{Motte}, {Andr{\'e}}, {Ward-Thompson} \&
  {Bontemps}}{{Motte} et~al.}{2001}]{2001A&A...372L..41M}
{Motte} F.,  {Andr{\'e}} P.,  {Ward-Thompson} D.,    {Bontemps} S.,  2001,
  A\&A, 372, L41

\bibitem[\protect\citeauthoryear{{Nutter} \& {Ward-Thompson}}{{Nutter} \&
  {Ward-Thompson}}{2007}]{2007MNRAS.374.1413N}
{Nutter} D.,  {Ward-Thompson} D.,  2007, MNRAS, 374, 1413

\bibitem[\protect\citeauthoryear{{Ossenkopf} \& {Henning}}{{Ossenkopf} \&
  {Henning}}{1994}]{1994A&A...291..943O}
{Ossenkopf} V.,  {Henning} T.,  1994, A\&A, 291, 943

\bibitem[\protect\citeauthoryear{{Papadopoulos} \& {Allen}}{{Papadopoulos} \&
  {Allen}}{2000}]{2000ApJ...537..631P}
{Papadopoulos} P.~P.,  {Allen} M.~L.,  2000, ApJ, 537, 631

\bibitem[\protect\citeauthoryear{{Reid} \& {Wilson}}{{Reid} \&
  {Wilson}}{2005}]{2005ApJ...625..891R}
{Reid} M.~A.,  {Wilson} C.~D.,  2005, ApJ, 625, 891

\bibitem[\protect\citeauthoryear{{Richer}, {Hills}, {Padman} \&
  {Russell}}{{Richer} et~al.}{1989}]{1989MNRAS.241..231R}
{Richer} J.~S.,  {Hills} R.~E.,  {Padman} R.,    {Russell} A.~P.~G.,  1989,
  MNRAS, 241, 231

\bibitem[\protect\citeauthoryear{{Seaquist}, {Yao}, {Dunne} \&
  {Cameron}}{{Seaquist} et~al.}{2004}]{2004MNRAS.349.1428S}
{Seaquist} E.,  {Yao} L.,  {Dunne} L.,    {Cameron} H.,  2004, MNRAS, 349, 1428

\bibitem[\protect\citeauthoryear{{Serabyn} \& {Weisstein}}{{Serabyn} \&
  {Weisstein}}{1995}]{1995ApJ...451..238S}
{Serabyn} E.,  {Weisstein} E.~W.,  1995, ApJ, 451, 238

\bibitem[\protect\citeauthoryear{{Shirley}, {Evans} II \& {Rawlings}}{{Shirley}
  et~al.}{2002}]{2002ApJ...575..337S}
{Shirley} Y.~L.,  {Evans} II N.~J.,    {Rawlings} J.~M.~C.,  2002, ApJ, 575,
  337

\bibitem[\protect\citeauthoryear{{Tothill}, {White}, {Matthews}, {McCutcheon},
  {McCaughrean} \& {Kenworthy}}{{Tothill} et~al.}{2002}]{2002ApJ...580..285T}
{Tothill} N.~F.~H.,  {White} G.~J.,  {Matthews} H.~E.,  {McCutcheon} W.~H.,
  {McCaughrean} M.~J.,    {Kenworthy} M.~A.,  2002, ApJ, 580, 285

\bibitem[\protect\citeauthoryear{{van Kempen}, {van Dishoeck}, {G{\"u}sten},
  {Kristensen}, {Schilke}, {Hogerheijde}, {Boland}, {Menten} \&
  {Wyrowski}}{{van Kempen} et~al.}{2009}]{2009A&A...507.1425V}
{van Kempen} T.~A.,  {van Dishoeck} E.~F.,  {G{\"u}sten} R.,  {Kristensen}
  L.~E.,  {Schilke} P.,  {Hogerheijde} M.~R.,  {Boland} W.,  {Menten} K.~M.,
  {Wyrowski} F.,  2009, A\&A, 507, 1425

\bibitem[\protect\citeauthoryear{{Ward-Thompson}, {Di Francesco}, {Hatchell},
  {Hogerheijde}, {Nutter}, {Bastien}, {Basu}, {Bonnell}, {Bowey} \&
  {Brunt}}{{Ward-Thompson} et~al.}{2007}]{2007PASP..119..855W}
{Ward-Thompson} D.,  {Di Francesco} J.,  {Hatchell} J.,  {Hogerheijde} M.~R.,
  {Nutter} D.,  {Bastien} P.,  {Basu} S.,  {Bonnell} I.,  {Bowey} J.,
  {Brunt} C.,  2007, PASP, 119, 855

\bibitem[\protect\citeauthoryear{{Ward-Thompson}, {Motte} \&
  {Andre}}{{Ward-Thompson} et~al.}{1999}]{1999MNRAS.305..143W}
{Ward-Thompson} D.,  {Motte} F.,    {Andre} P.,  1999, MNRAS, 305, 143

\bibitem[\protect\citeauthoryear{{Wilking}, {Meyer}, {Greene}, {Mikhail} \&
  {Carlson}}{{Wilking} et~al.}{2004}]{2004AJ....127.1131W}
{Wilking} B.~A.,  {Meyer} M.~R.,  {Greene} T.~P.,  {Mikhail} A.,    {Carlson}
  G.,  2004, AJ, 127, 1131

\bibitem[\protect\citeauthoryear{{Wilson}}{{Wilson}}{2009}]{2009arXiv0903.0562W}
{Wilson} T.~L.,  2009, ArXiv e-prints

\bibitem[\protect\citeauthoryear{{Y{\i}ld{\i}z}, {van Dishoeck}, {Kristensen},
  {Visser}, {J{\o}rgensen}, {Herczeg}, {van Kempen}, {Hogerheijde} \&
  {Doty}}{{Y{\i}ld{\i}z} et~al.}{2010}]{2010A&A...521L..40Y}
{Y{\i}ld{\i}z} U.~A.,  {van Dishoeck} E.~F.,  {Kristensen} L.~E.,  {Visser} R.,
   {J{\o}rgensen} J.~K.,  {Herczeg} G.~J.,  {van Kempen} T.~A.,  {Hogerheijde}
  M.~R.,    {Doty} S.~D.,  2010, A\&A, 521, L40+

\bibitem[\protect\citeauthoryear{{Young}, {Shirley}, {Evans} II \&
  {Rawlings}}{{Young} et~al.}{2003}]{2003ApJS..145..111Y}
{Young} C.~H.,  {Shirley} Y.~L.,  {Evans} II N.~J.,    {Rawlings} J.~M.~C.,
  2003, ApJS, 145, 111

\bibitem[\protect\citeauthoryear{{Zhu}, {Seaquist} \& {Kuno}}{{Zhu}
  et~al.}{2003}]{2003ApJ...588..243Z}
{Zhu} M.,  {Seaquist} E.~R.,    {Kuno} N.,  2003, ApJ, 588, 243

\end{thebibliography}
%\pagebreak
%\tablefirsthead{ \multicolumn{1}{l}{Region} & \multicolumn{1}{c}{Source} & RA & Dec & Other Source Number \\ & & \multicolumn{1}{c}{(J2000)} & (J2000) & \\ \hline}
%\tablehead{\multicolumn{5}{l}{\small continued from previous page} \\ \hline \ \multicolumn{1}{l}{Region} & \multicolumn{1}{c}{Source} & RA & Dec & Other Source Number \\ & & \multicolumn{1}{c}{(J2000)} & (J2000) & \\ \hline}
%\tablelasttail{\hline}
%\topcaption{List of sources used for the study of $^{12}$CO contamination to the SCUBA-2 850~$\micron$ dust continuum.}
%\begin{supertabular*}{l c c c c }
%\hline
%\end{supertabular*}
%\end{center}
%\end{minipage}
\onecolumn
\begin{center}
\begin{longtable}{l | c | cc | c}
%\begin{longtable}{p{0.2\columnwidth} p{0.12\columnwidth} p{0.12\columnwidth} p{0.12\columnwidth} p{0.12\columnwidth}}

\caption{List of sources used for the study of $^{12}$CO contamination to the SCUBA-2 850~$\micron$ dust continuum.  Source numbers correspond to the arbitrary number assigned to sources for this study.  Source numbers corresponding to original studies (NGC~1333 sources obtained from \citealt{2007A&A...468.1009H} and NGC~2071 and NGC~2024 sources obtained from \citealt{2007MNRAS.374.1413N}) are listed under `Other Source ID'.}\\
%\multicolumn{1}Region & Source & RA & Dec & Other Source Number \\ & & (J2000) & (J2000) & \\ \hline
%\begin{longtable*}{p{0.2\columnwidth}  p{0.12\columnwidth}  p{0.12\columnwidth} p{0.12\columnwidth} p{0.12\columnwidth}}
\hline \hline Region & Source & RA & Dec & Other Source ID  \\ & & (J2000) & (J2000) & \\  \hline \endfirsthead
\caption{continued from previous page.}\\
%\multicolumn{5}{l}{\small \em{continued from previous page}} \\ 
\hline \hline Region & Source & RA & Dec & Other Source ID \\ & & (J2000) & (J2000) & \\  \hline \endhead
NGC~1333 & 1 & 03:29:10.4 & 31:13:30 & HRF41\\
& 2 & 03:29:12.0 & 31:13:10 & HRF42\\
& 3 & 03:29:03.2 & 31:15:59 & HRF43\\
& 6 & 03:28:55.3 & 31:14:36 & HRF44\\
& 13 & 03:29:01.4 & 31:20:29 & HRF45\\
& 14 & 03:29:11.0 & 31:18:27 & HRF46\\
& 17 & 03:28:59.7 & 31:21:34 & HRF47\\
& 18 & 03:29:13.6 & 31:13:55 & HRF48\\
& 20 & 03:28:36.7 & 31:13:30 & HRF49\\
& 21 & 03:29:06.5 & 31:15:39 & HRF50\\
& 22 & 03:29:08.8 & 31:15:18 & HRF51\\
& 24 & 03:29:03.7 & 31:14:53 & HRF52\\
& 26 & 03:29:04.5 & 31:20:59 & HRF53\\
& 29 & 03:29:10.7 & 31:21:45 & HRF54\\
& 30 & 03:28:40.4 & 31:17:51 & HRF55\\
& 32 & 03:29:07.7 & 31:21:57 & HRF56\\
& 33 & 03:29:18.2 & 31:25:11 & HRF57\\
& 35 & 03:29:16.5 & 31:12:35 & HRF59\\
& 38 & 03:28:39.4 & 31:18:27 & HRF60\\
& 39 & 03:29:17.3 & 31:27:50 & HRF61\\
& 42 & 03:29:07.1 & 31:17:24 & HRF62\\
& 43 & 03:29:18.8 & 31:23:17 & HRF63\\
& 46 & 03:29:25.5 & 31:28:18 & HRF64\\
& 47 & 03:29:00.4 & 31:12:02 & HRF65\\
& 53 & 03:29:05.3 & 31:22:11 & HRF66\\
& 55 & 03:29:19.7 & 31:23:56 & HRF67\\
& 62 & 03:28:56.2 & 31:19:13 & HRF68\\
& 63 & 03:28:34.4 & 31:06:59 & HRF69\\
& 66 & 03:29:15.3 & 31:20:31 & HRF70\\
& 73 & 03:28:38.7 & 31:05:57 & HRF71\\
& 77 & 03:29:19.1 & 31:11:38 & HRF72\\
& 81 & 03:28:32.5 & 31:11:08 & HRF74\\
& 83 & 03:28:42.6 & 31:06:10 & HRF75\\
& 89 & 03:29:04.9 & 31:18:41 & Bolo44\\
& 94 & 03:28:32.7 & 31:04:56 & Bolo26\\ \hline
NGC~2071 & 51 & 05:47:23.7 & 00:11:02 & BN-547237+01102 \\
& 52 & 05:47:06.8 & 00:12:30 & BN-547068+01230 \\
& 53 & 05:47:10.6 & 00:13:18 & BN-547106+01318\\
& 54 & 05:47:05.1 & 00:13:21 & BN-547051+01321\\
& 55 & 05:47:05.0 & 00:14:49 & BN-547050+01449\\
& 56 & 05:47:23.9 & 00:15:07 & BN-547239+01507\\
& 57 & 05:47:12.4 & 00:15:37 & BN-547124+01537\\
& 58 & 05:47:10.4 & 00:15:53 & BN-547104+01553\\
& 59 & 05:47:19.9 & 00:16:03 & BN-547199+01603\\
& 60 & 05:47:04.8 & 00:17:07 & BN-547048+01707\\
& 61 & 05:47:01.5 & 00:17:55 & BN-547015+01755\\
& 62 & 05:47:08.7 & 00:18:17 & BN-547087+01817\\
& 63 & 05:47:15.2 & 00:18:30 & BN-547152+01830\\
& 64 & 05:47:25.3 & 00:18:48 & BN-547253+01848\\
& 65 & 05:47:33.6 & 00:19:02 & BN-547336+01902\\
& 66 & 05:46:28.3 & 00:19:28 & BN-546283+01928\\
& 67 & 05:47:03.4 & 00:19:50 & BN-547034+01950\\
& 68 & 05:47:26.7 & 00:19:53 & BN-547267+01953\\
& 69 & 05:47:37.7 & 00:20:01 & BN-547377+02001\\
& 70 & 05:46:57.6 & 00:20:09 & BN-546576+02009\\
& 71 & 05:46:29.4 & 00:20:10 & BN-546294+02010\\
& 72 & 05:47:34.9 & 00:20:20 & BN-547349+02020\\
& 73 & 05:47:32.5 & 00:20:26 & BN-547325+02026\\
& 74 & 05:47:01.0 & 00:20:42 & BN-547010+02042\\
& 75 & 05:47:25.2 & 00:20:59 & BN-547252+02059\\
& 76 & 05:47:10.3 & 00:21:12 & BN-547103+02112\\
& 77 & 05:46:28.7 & 00:21:14 & BN-546287+02114\\
& 78 & 05:47:16.0 & 00:21:23 & BN-547160+02123\\
& 79 & 05:47:04.1 & 00:21:58 & BN-547041+02158\\
& 80 & 05:46:25.3 & 00:22:20 & BN-546253+02220\\
& 81 & 05:46:52.8 & 00:22:23 & BN-546528+02223\\
& 82 & 05:47:11.9 & 00:22:23 & BN-547119+02223\\
& 83 & 05:47:19.7 & 00:22:31 & BN-547197+02231\\
& 84 & 05:47:06.9 & 00:22:39 & BN-547069+02239\\
& 85 & 05:47:17.5 & 00:22:40 & BN-547175+02240\\
& 86 & 05:46:59.1 & 00:22:59 & BN-546591+02259\\
& 87 & 05:47:12.4 & 00:23:11 & BN-547124+02311\\
& 88 & 05:47:06.7 & 00:23:14 & BN-547067+02314\\
& 89 & 05:46:54.7 & 00:23:24 & BN-546547+02324\\
& 90 & 05:47:10.4 & 00:23:27 & BN-547104+02327\\
& 91 & 05:47:08.9 & 00:23:56 & BN-547089+02356\\
& 92 & 05:46:57.2 & 00:23:56 & BN-546572+02356\\
& 93 & 05:46:34.7 & 00:23:59 & BN-546347+02359\\
& 94 & 05:46:58.0 & 00:24:26 & BN-546580+02426\\
& 95 & 05:47:01.7 & 00:24:52 & BN-547017+02452\\
& 96 & 05:46:25.7 & 00:24:56 & BN-546257+02456\\
& 97 & 05:47:08.0 & 00:25:05 & BN-547080+02505\\
& 98 & 05:46:45.9 & 00:25:07 & BN-546459+02507\\
& 99 & 05:47:01.4 & 00:26:14 & BN-547014+02614\\
& 100 & 05:46:38.0 & 00:26:53 & BN-546380+02653\\ \hline
NGC~2024 & 33 & 05 42 03.0 & -02 04 23 & BS-542030-20423 \\
& 34 & 05 42 10.3 & -02 04 20 & BS-542103-20420\\
& 35 & 05 42 03.5 & -02 02 24 & BS-542035-20224\\
& 36 & 05 41 57.1 & -02 01 00 & BS-541571-20100\\
& 37 & 05 41 52.9 & -02 00 21 & BS-541529-20021\\
& 38 & 05 41 49.3 & -01 59 38 & BS-541493-15938\\
& 40 & 05 42 00.0 & -01 58 01 & BS-542000-15801\\
& 41 & 05 41 49.1 & -01 58 03 & BS-541491-15803\\
& 42 & 05 41 45.2 & -01 56 31 & BS-541452-15631\\
& 43 & 05 41 35.4 & -01 56 29 & BS-541354-15629\\
& 45 & 05 41 44.5 & -01 55 39 & BS-541445-15539\\
& 46 & 05 41 44.2 & -01 54 43 & BS-541442-15443\\
& 47 & 05 41 32.1 & -01 54 26 & BS-541321-15426\\
& 48 & 05 41 19.9 & -01 54 16 & BS-541199-15416\\
&  49 & 05 41 42.0 & -01 53 59 & BS-541420-15359\\
& 51 & 05 41 44.2 & -01 52 41 & BS-541442-15241\\
& 52 & 05 41 16.6 & -01 51 19 & BS-541166-15119\\
& 53 & 05 41 36.7 & -01 51 06 & BS-541367-15106\\
& 54 & 05 41 23.4 & -01 50 27 & BS-541234-15027\\
& 55 & 05 41 32.9 & -01 49 53 & BS-541329-14953\\
& 56 & 05 41 36.4 & -01 49 24 & BS-541364-14924\\
& 57 & 05 41 27.6 & -01 48 13 & BS-541276-14813\\
& 58 & 05 41 11.3 & -01 48 12 & BS-541113-14812\\
& 59 & 05 41 13.3 & -01 47 35 & BS-541133-14735\\\hline
\label{table:source_list}
\end{longtable}
\twocolumn
\end{center}

\appendix

\section{Molecular Outflow Spectra Criteria}
The full version of this Appendix is available as Supporting Information to the online version of this article.

\begin{figure*}
\includegraphics[width=6.5in]{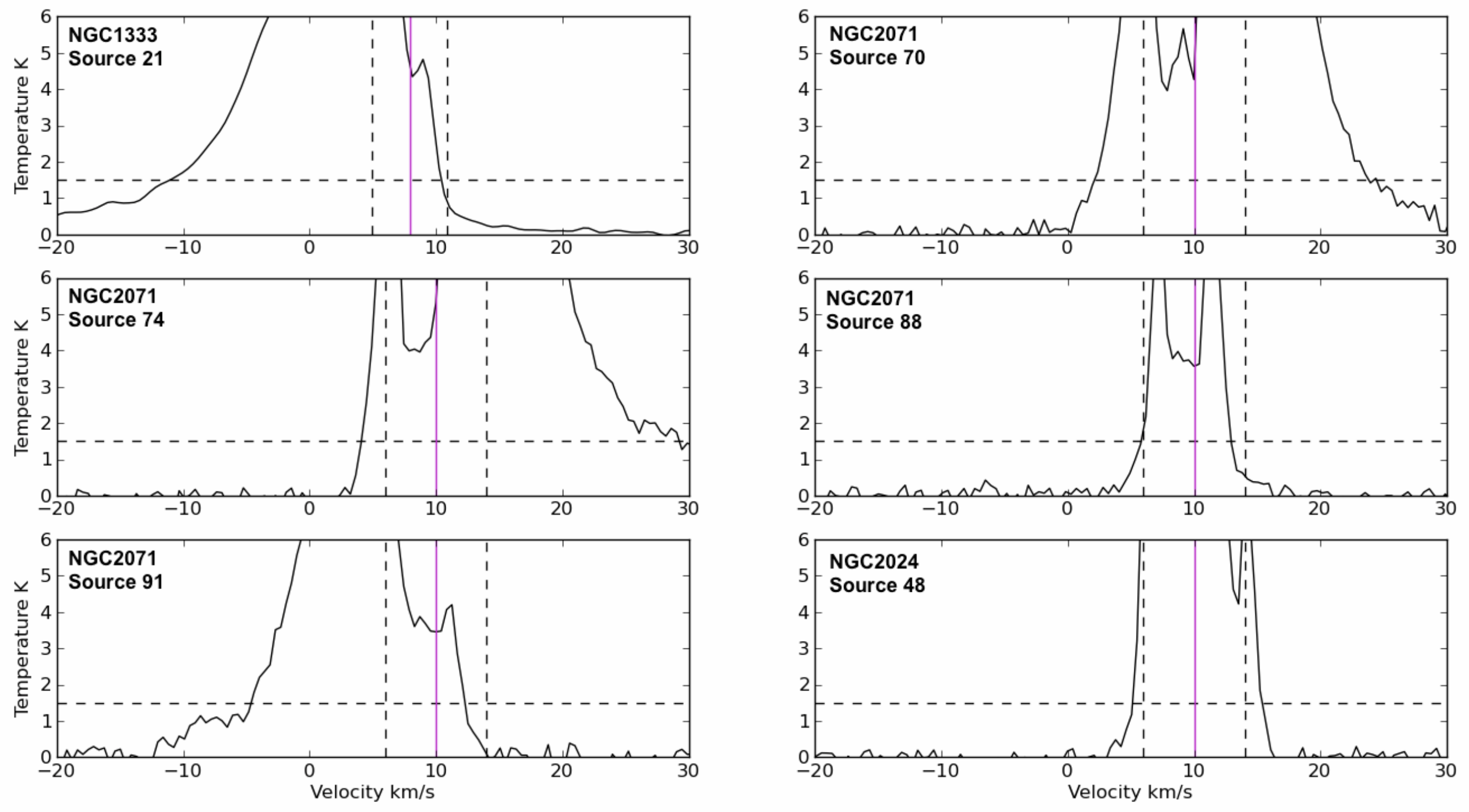}
\caption{$^{12}$CO~3--2 spectra for sources further analysed in Section~\ref{outflows} regarding the molecular outflow analysis.  The core velocities are listed in Section~\ref{ngc_1333_flows} for NGC~1333 and in Table~\ref{moloutflow} for NGC~2071 and NGC~2024.  Parameters $\pm$3~km~s$^{-1}$ for NGC~1333 and $\pm$4~km~s$^{-1}$ for NGC~2071 and NGC~2024 at 1.5~K were used to classify the presence of outflows. }
\label{fig:spectra_sources}
\end{figure*}

\end{document}